\documentclass[letter,12pt]{article}
\usepackage{epsfig,amsmath,amssymb,verbatim,mathrsfs}
\usepackage{rotating}
\usepackage{cite}
\usepackage{verbatim}
 \pdfoutput=1
\def\g{\gamma}
\def\beq{\begin{equation}}
\def\eeq{\end{equation}}
\def\bea{\begin{eqnarray}}
\def\eea{\end{eqnarray}}
\def\bmat{\begin{pmatrix}}
\def\emat{\end{pmatrix}}
\def\bei{\begin{itemize}}
\def\eei{\end{itemize}}
\def\Im{\, {\rm Im}}

\begin{document}
\baselineskip=18pt

\begin{titlepage}

\noindent
\begin{flushright}
CERN-PH-TH/2011-284\\
MCTP-11-40
\end{flushright}
\vspace{1cm}

\begin{center}
  \begin{Large}
    \begin{bf}
   Probing Quartic Neutral Gauge Boson Couplings\\  
   using diffractive photon fusion at the LHC

        \end{bf}
  \end{Large}
\end{center}

\vspace{0.5cm}
\begin{center}
\begin{large}

Rick S. Gupta \\

\end{large}

\vspace{0.3cm}
\begin{it}
CERN, Theoretical Physics, CH-1211 Geneva 23, Switzerland \\
\vspace{0.1cm}
and\\

Michigan Center for Theoretical Physics (MCTP), \\
        ~~University of Michigan, Ann Arbor, MI 48109-1120, USA \\
\end{it}

\vspace{1cm}
\end{center}

\begin{abstract}
\noindent
A complete list of operators contributing  at the lowest order to Quartic Neutral Gauge Boson Couplings involving photons and $Z$-bosons,  is presented. We show that, for the couplings we consider,  the lowest order contribution is from dimension 8 operators in the case when a light Higgs is present and from dimension 6 operators in the higgsless case where electroweak symmetry is non-linearly realized. We also show that these operators  are generated by exchange of the Kaluza-Klein partners of the graviton in extra-dimensional models. We then explore the possibility of probing these couplings in the diffractive photon fusion processes $pp(\g\g \to \g\g)pp$ and $pp(\g\g \to ZZ)pp$ at the 14 TeV LHC. We find that the $\g\g\g\g$ coupling can be probed most sensitively and values as small as $1/(1.8$ TeV)$^{4}$ can be measured. For the $\g\g ZZ$ coupling, values as small as $1/(850$ GeV)$^{4}$ and $1/(1.9$ TeV)$^2$ can be probed  in the light Higgs and higgsless cases respectively,  which is an improvement by orders of magnitude over  existing limits. 
\end{abstract}

\vspace{2cm}

\begin{flushleft}
\begin{small}
November 2011
\end{small}
\end{flushleft}

\end{titlepage}
\tableofcontents
\newpage

\section{Introduction}
 
The Standard Model (SM) has been tested very accurately by experiments. There are, however, many theoretical reasons to believe that there is physics beyond the SM. Some of these motivations, like the hierarchy problem and the existence of dark matter, point to the existence of new physics at the TeV scale. LEP-2 precision data  and flavor constraints seem to favor a scenario with a mass gap between a light Higgs ($m_H\lesssim 200$ GeV) and new physics at the scale of a few TeV.  A model independent way of parametrizing the effects of new physics in such a scenario is to use the effective field theory approach. All possible operators allowed by the symmetries of the theory are included, suppressed by appropriate powers of the  cut-off $\Lambda$. If $\Lambda$ is the order of a few TeV, these operators can be directly measured at the Large Hadron Collider (LHC).

These operators are expected to give rise to anomalous triple~\cite{Buchmuller:1985jz, tgc} and quartic gauge couplings~\cite{qgc}. In this work we discuss a special class of these couplings:  the Quartic Neutral Gauge Boson Couplings (QNGC), that is, quartic vertices involving only the neutral gauge bosons, $\gamma$ and $Z$. QNGCs are special because as we will show  they do not exist in the SM  and receive their lowest order contributions from dimension 8 operators.  Thus the measurement of these couplings would indicate directly the presence of dimension 8 operators\footnote{In the case of the $\g\g ZZ$ coupling there is a non-local contribution from the $\g\g \to h^* \to ZZ$ process which is of an  order lower than dimension 8 contributions. However, as we discuss later, this contribution can be subtracted if the $h\to \g \g$ partial width is known.} in a scenario where a light Higgs is present.   This is not true, for example, in the case of $\g\g W^+W^-$ and $ZZW^+W^-$ couplings which get contributions from the SM lagrangian and its dimension 6 extension~\cite{Buchmuller:1985jz}. Thus QNGCs can be very useful in probing new physics scenarios with a light Higgs that exclusively generate  dimension 8 operators. One such example that generates only dimension 8 operators at tree-level is the exchange of the spin 2 Kaluza-Klein excitations of the graviton in models with large extra dimensions. We will see how integrating out these massive modes generates QNGCs and how probing these couplings would allow us to probe the fundamental Planck scale in these extra dimensional theories. 

We also consider the higgsless case where electroweak symmetry is non-linearly realized. In this case, with the exception of the $ZZZZ$ coupling, QNGCs do not appear at the dimension 4 level and the lowest order contribution comes from dimension 6 operators. This is unlike quartic gauge couplings having $W^{+/-}$ bosons which always appear first at the dimension 4 level. Thus  in this case also, unlike processes involving quartic gauge couplings with $W^{+/-}$ bosons, processes involving QNGCs can directly probe higher order operators (in this case dimension 6 operators).

\begin{figure}[t]
\begin{center}
\includegraphics[width=0.65\columnwidth]{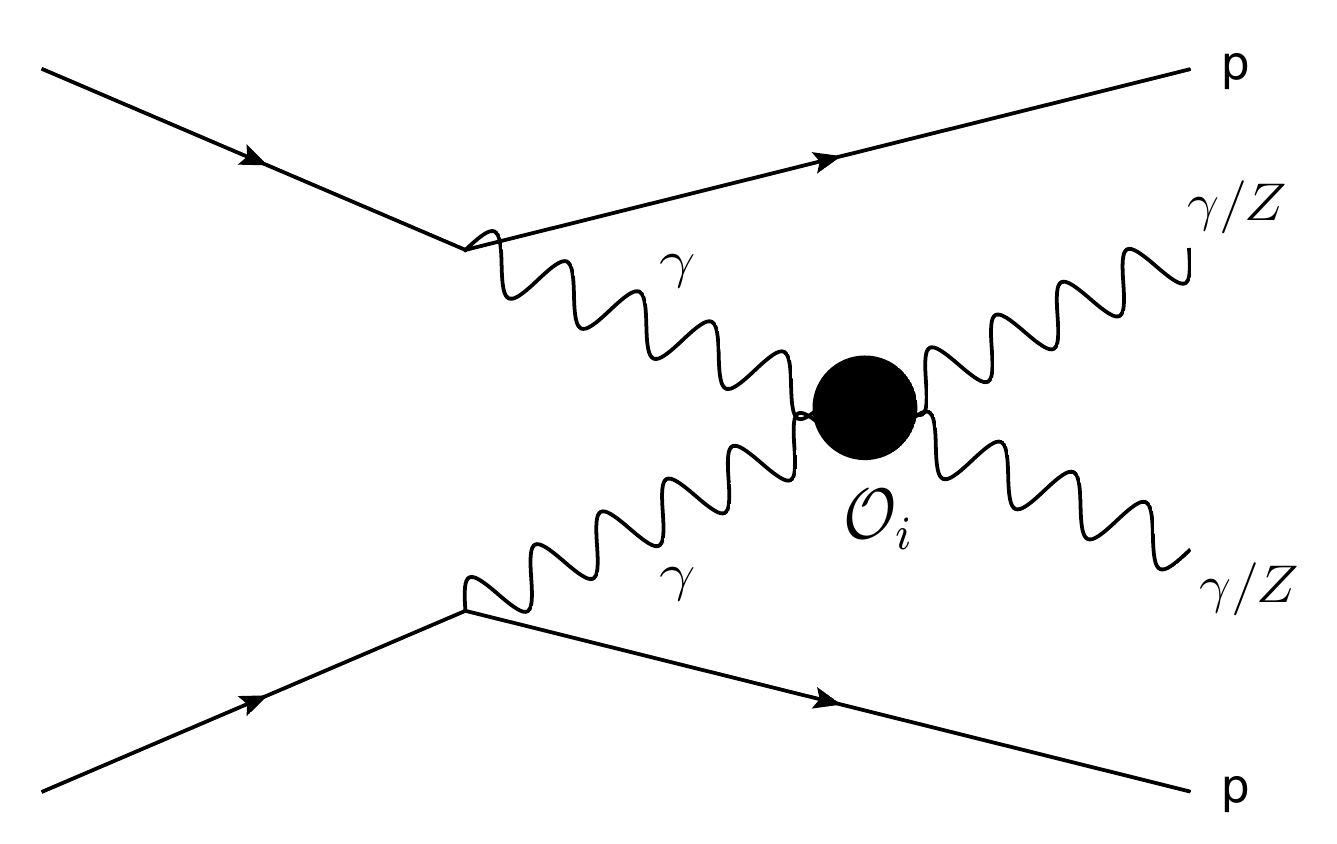}
\end{center}
\caption{The diffractive photon fusion processes $pp(\g\g \to \g\g)pp$ and $pp(\g\g \to ZZ)pp$. The outgoing protons can be detected by very forward detectors to be installed by ATLAS and CMS. In the figure above ${\cal O}_i$ represents operators contributing to Quartic Neutral Gauge Couplings (QNGCs).}
\label{fd}
\end{figure}

In this work we will explore the possibility of measuring the $\g\g  \g\g$ and $\g\g  ZZ$ couplings in the diffractive photon fusion processes,  $pp(\g\g \to \g\g)pp$ and $pp(\g\g \to ZZ)pp$ (see Fig.~\ref{fd}) respectively. There are plans to install very forward detectors by the  ATLAS and CMS collaborations~\cite{AFP} which can detect  protons that scatter diffractively at small angles and thus can identify such processes.  To the best of our knowledge, this is the first study on the LHC sensitivity of  the measurement of the $\g\g\g\g$ coupling. There have been previous studies for the  $\g\g ZZ$ coupling,  but these studies focussed  only on the higgsless case. In Refs.~\cite{Dervan:1999as, Eboli:2000ad} probing the $\g\g ZZ$ coupling by inelastic processes like $pp \to\g\g\g$, $pp \to \g \g Z$ and $pp \to jj (ZZ \to \g\g)\to jj \g\g$ has been studied, whereas, in Refs.~\cite{Pierzchala:2008xc,Kepka:2008yx} measurement of this coupling in the diffractive process $pp(\g\g \to ZZ)pp$, that we will  study in this work too,   has been explored. In this work, however, we  consider both the light Higgs and the higgsless cases. As we will  see, in the higgsless case considered in the previous studies only a subset of all the operators relevant to the light Higgs case are important. 

Let us now see what are the advantages of  diffractive photon fusion processes in measuring these couplings. Even if a process can be traced back to a definite set of operators as is the case here, it is rarely the case that a particular collider signature can be  traced back to a unique process. For this reason many different, complementary measurements are usually required to uncover the underlying new physics processes. For example consider the inelastic counterpart of the signature we are considering for the $\g\g ZZ$ coupling, the  $pp \to jj (\g \g\to ZZ) \to jj ZZ$ process or the similar vector boson fusion (VBF) process $pp \to jj (ZZ \to \g\g)\to jj \g\g$. Although these signatures would have a much larger cross-section than the diffractive signature we are considering, if an excess is observed in the $jjZZ$ or $jj\g\g$ final states it would be hard to reconstruct the exact process responsible for it because of the many different new physics processes in addition to QNGCs that can have this signature\footnote{The $pp \to jj (\g\g \to ZZ)\to jj ZZ$  process is experimentally challenging for a separate reason too which is that  the two jets would not have the special properties of VBF jets.   VBF jets have a large rapidity gap between them and have high $p_T$ (see for example pg. 1271-1305 of~\cite{tdr}). For reasons mentioned in Section~\ref{epha} the $p_T$  of the  jets in the process $pp \to jj (\g\g \to ZZ)\to jj ZZ$ is approximately equal to the photon virtuality and thus expected to be very small. So while the photon fusion jets would have a large rapidity gap too, they will have very low $p_T$}. The triple gauge boson production processes $pp \to \g \g Z$ (studied previously in Ref.~\cite{Dervan:1999as, Eboli:2000ad}) and $pp \to\g\g\g$ are somewhat better in this respect but, again, because the intermediate state in  $pp \to\g\g Z (\g\g \g)$ cannot be known,  it would not be possible to conclude with certainty that QNGCs are responsible, if an excess is seen.  The diffractive signals we study in this work are interesting because exclusive final states where two protons have been detected in the forward detectors can arise only from diffractive photon fusion or exclusive pomeron fusion. In the latter case the underlying subprocess would be $gg\to \g\g/ ZZ$. Thus the inverse problem of pinning down the new physics responsible for an excess in the $pp\g\g$ and $ppZZ$ final states is relatively less ambiguous as there are only two new physics possibilities namely the enhancement of the $\g\g \to \g \g /ZZ$ processes and/or the enhancement of the $gg \to \g\g /ZZ$ processes. As we will discuss later, exclusive pomeron fusion  processes are, however, expected to have a much smaller cross-section as compared to photon fusion processes.
 
\section {Operators that give rise to Quartic Neutral Gauge Boson Couplings}

\subsection{Light Higgs case}
\label{lh}
We want to write down the lowest order contribution form higher dimensional operators to  QNGCs, that is, quartic vertices involving only the neutral gauge bosons $\gamma$ and $Z$. We will consider only CP conserving operators here hence we will not use the dual field strength tensors like $\tilde{B}_{\mu\nu}$. In gauge invariant operators the electroweak vector boson fields can appear either from the field strengths $B_{\mu\nu}$ and $W^I_{\mu\nu}$ or  covariant derivatives acting on the Higgs field, that is  factors like,
\beq
D_\mu\Phi=\partial_\mu \Phi -i g\frac{\tau_I}{2}W^I_\mu \Phi -i \frac{g'}{2}  B_\mu \Phi,
\eeq
 provided the Higgs field, $\Phi$, gets a vacuum expectation value (VEV). In the above equation $\tau_I$ are the Pauli matrices. Note that the $\epsilon_{IJK}W_J W_K$ part of $W^I_{\mu\nu}$ cannot contribute to quartic neutral gauge boson couplings as we must have $I=J=3$ to get photons or $Z$ fields unlike for instance in the case of the $\gamma \gamma W^+W^-$ coupling, where two gauge boson fields can come from the same  field strength tensor $W^I_{\mu\nu}$. This has the important implication that the lowest order contribution to quartic neutral gauge boson couplings comes from dimension 8 operators\footnote{This fact is also true for $gg \g\g$ and $gg ZZ$ couplings ($g$ being the gluon).} because we need either a field strength or $D_\mu\Phi$ factor, both dimension 2 operators, for each of the four  gauge bosons.  There must be an even number of $D_\mu\Phi$ factors along with the field strength tensors in these operators  because otherwise  there are an odd number of Lorentz indices in total and it is impossible to contract all of them. Thus we see that the operators we are interested in  can have either  four covariant derivatives, two field strengths and two covariant derivatives  or four field strengths. We thus obtain the following lagrangian of dimension 8 operators,
\bea
{\cal L}_{QNGC} &=& \frac{c_1}{\Lambda^4}D_\mu\Phi^\dagger D^\mu\Phi D_\nu\Phi^\dagger D^\nu \Phi+\frac{c_2}{\Lambda^4}D_\mu\Phi^\dagger D_\nu\Phi D^\mu\Phi^\dagger D^\nu \Phi+\frac{c_3}{\Lambda^4}D_\rho\Phi^\dagger D^\rho\Phi B_{\mu \nu}B^{\mu \nu}
\nonumber\\
&& +\frac{c_4}{\Lambda^4}D_\rho\Phi^\dagger D^\rho\Phi W^I_{\mu \nu}W^{I\mu \nu}+\frac{c_5}{\Lambda^4}D_\rho\Phi^\dagger \sigma^I D^\rho\Phi B_{\mu \nu}W^{I\mu \nu}+\frac{c_6}{\Lambda^4}D_\rho\Phi^\dagger 
D^\nu\Phi B_{\mu \nu}B^{\mu \rho}
\nonumber\\
&&+\frac{c_7}{\Lambda^4}D_\rho\Phi^\dagger D^\nu\Phi W^I_{\mu \nu}W^{I \mu\rho}+\frac{c_8}{\Lambda^4}B_{\rho \sigma}B^{\rho \sigma} B_{\mu \nu}B^{\mu \nu}+\frac{c_{9}}{\Lambda^4}W^I_{\rho \sigma}W^{I\rho \sigma} W^J_{\mu \nu}W^{J\mu \nu}
\nonumber\\&&+\frac{c_{10}}{\Lambda^4}W^I_{\rho \sigma}W^{J\rho \sigma} W^I_{\mu \nu}W^{J\mu \nu}+\frac{c_{11}}{\Lambda^4}B_{\rho \sigma}B^{\rho \sigma} W^I_{\mu \nu}W^{I\mu \nu}+\frac{c_{12}}{\Lambda^4}B_{\rho \sigma}W^{I\rho \sigma} B_{\mu \nu}W^{I\mu \nu}
\nonumber\\
&& +
\frac{c_{13}}{\Lambda^4}B_{\rho \sigma}B^{\sigma \nu} B_{\mu \nu}B^{\mu \rho}
+\frac{c_{14}}{\Lambda^4}W^I_{\rho \sigma}W^{I\sigma \nu} W^J_{\mu \nu}W^{J\mu \rho}+\frac{c_{15}}{\Lambda^4}W^I_{\rho \sigma}W^{J\sigma \nu} W^I_{\mu \nu}W^{J\mu \rho}\nonumber\\
&&+\frac{c_{16}}{\Lambda^4}B_{\rho \sigma}B^{\sigma \nu} W^I_{\mu \nu}W^{I\mu \rho}+\frac{c_{17}}{\Lambda^4}B_{\rho \sigma}W^{I\sigma \nu} B_{\mu \nu}W^{I\mu \rho}.
\label{L8}
\eea
Note that, in the above list,  the operators $B_{\mu \nu} \epsilon_{IJK}W^{I\mu\nu}W^J_{\rho\sigma}W^{K\rho\sigma}$,  $B_{\mu \nu} \epsilon_{IJK}(\Phi^\dagger \sigma^I \Phi) W^J_{\rho\sigma}W^{K\rho\sigma}$ and \textrm{$B_{\mu \nu} \epsilon_{IJK}W^{I\mu\rho}W^{J\nu\sigma}W^K_{\rho\sigma}$} are absent because they are all equal to zero. In the first two cases $\epsilon_{IJK}$ is antisymmetric in $J$ and $K$ whereas the rest of the operator is symmetric in $J$ and $K$ and in the third case $B_{\mu \nu}$ is antisymmetric in $\mu$ and $\nu$ whereas the rest of the operator is symmetric in  $\mu$ and $\nu$.  Operators with two $\sigma^I$s do not appear above as these can be reduced to operators in our list using the identity,  $\sigma^I\sigma^J =\delta^{IJ}+i \epsilon^{IJK}\sigma_K$. Also notice that there are no operators like $\Phi^\dagger D^\mu D_\nu \Phi B_{\mu\rho}B^{\nu \rho}$. This is because such operators can be expressed as linear combinations of total derivatives, operators already in the list and operators that do not contribute to QNGCs, as follows,
\bea
\Phi^\dagger D^\mu D_\nu \Phi B_{\mu\rho}B^{\nu \rho} &=& \partial^\mu(\Phi^\dagger  D_\nu \Phi B_{\mu\rho}B^{\nu \rho})-D^\mu\Phi^\dagger  D_\nu \Phi B_{\mu\rho}B^{\nu \rho} \nonumber\\
&&-\Phi^\dagger  D_\nu \Phi \partial^\mu(B_{\mu\rho}B^{\nu \rho})
\eea
where we have used $\partial_\mu (\Phi^\dagger D_\nu \Phi) =D_\mu \Phi^\dagger D_\nu \Phi +\Phi^\dagger(D_\mu D_\nu \Phi) $. Finally, operators with two Levi Civita tensors,like $\epsilon_{\mu\nu\rho\sigma}\epsilon_{\alpha\beta\gamma\delta}B^{\mu\alpha}B^{\nu \beta}B^{\rho \gamma}B^{\sigma\delta}$,  which corresponds to taking two dual field strength tensors,   are not in the list. This is because using the identity,
\beq
\epsilon_{i_1 i_2 i_3 i_4}\epsilon_{j_1 j_2 j_3 j_4} =24~g_{j_i k_i}g_{j_2 k_2}g_{j_3 k_3}g_{j_4 k_4} \delta^{k_1}_{[i_i}.... \delta^{k_4}_{i_4]}
\eeq
we can express such operators in terms of operators contracted using metric tensors which are already in our list. 

If we rewrite these operators in terms of the fields $A$ and $Z$ defined by,
\bea
\left( \begin{array}{c}
B\\W_3 
\end{array}\right)=\left( \begin{array}{cc}
c_w& -s_w \\
 s_w &c_w
 \end{array}\right)\left( \begin{array}{c}
A\\Z 
\end{array}\right)
\label{transform1}
\eea
and the field strengths,
\bea
F_{\mu\nu}&=&\partial_\mu A_\nu-\partial_\nu A_\mu\nonumber\\
Z_{\mu\nu}&=&\partial_\mu Z_\nu-\partial_\nu Z_\mu.
\label{transform2}
\eea
we will get  $ZZZZ$, $\gamma ZZZ$, $\gamma \gamma ZZ$, $\gamma \gamma \gamma Z$ and $\gamma \gamma \gamma \gamma$ couplings. In this work we will explore the possibility of measuring these vertices by diffractive processes involving photon exchange, that is processes like $pp(\g\g\to X)pp$. Such processes can measure the  $\gamma \gamma ZZ$, $\gamma \gamma \gamma Z$ and $\gamma \gamma \gamma \gamma$ vertices but we will focus only  on the  $\gamma \gamma ZZ$ and $\gamma \gamma \gamma \gamma$ vertices here (we give the complete list of the $\g\g\g Z$, $\gamma ZZZ$ and $ZZZZ$ couplings in Appendix A). Expressing the operators above in terms of $A$ and $Z$ fields and the respective field strengths using eq.~\ref{transform1},~\ref{transform2} we  get,
\bea
{\cal L}^{\g\g\g\g}_{QNGC} &=&  \frac{a_1^{\g\g}}{\Lambda^4} F_{\mu \nu}F^{\mu \nu} F_{\rho \sigma}F^{\rho \sigma} +\frac{a_2^{\g\g}}{\Lambda^4} F_{\mu \nu}F^{\mu \rho} F_{\rho \sigma}F^{\sigma \nu}
\nonumber\\
{\cal L}^{\g\g ZZ}_{QNGC} &=& \frac{a_1^{ZZ}}{\Lambda^4}\frac{M_Z^2}{2} F_{\mu \nu}F^{\mu \nu} Z_\rho Z^\rho+ \frac{a_2^{ZZ}}{\Lambda^4}\frac{M_Z^2}{2}F_{\mu \nu}F^{\mu \rho} Z_\rho Z^\nu+ \frac{a_3^{ZZ}}{\Lambda^4} F_{\mu \nu}F^{\mu \nu} Z_{\rho \sigma}Z^{\rho \sigma}
\nonumber\\&&
+ \frac{a_4^{ZZ}}{\Lambda^4} F_{\mu \nu}Z^{\mu \nu} F_{\rho \sigma}Z^{\rho \sigma}
+ \frac{a_5^{ZZ}}{\Lambda^4} F_{\mu \nu}F^{\mu \rho} Z_{\rho \sigma}Z^{\sigma \nu}+\frac{a_6^{ZZ}}{\Lambda^4} F_{\mu \nu}Z^{\mu \rho} F_{\rho \sigma}Z^{\sigma \nu}
\label{Lph}
\eea
where,
\bea
a_1^{\g\g}&=&c_w^4 c_{8}+ s_w^4 c_{9}+c_w^2 s_w^2 (c_{10}+c_{11})\nonumber\\
a_2^{\g\g}&=&c_w^4 c_{13}+ s_w^4 c_{14}+c_w^2 s_w^2 (c_{15}+c_{16})\nonumber\\
a_1^{ZZ}&=&c_w^2 c_3 +s_w^2 c_4-c_w s_w c_5\nonumber\\
a_2^{ZZ}&=&c_w^2 c_6 +s_w^2 c_7\nonumber\\
a_3^{ZZ}&=&2 c_w^2 s_w^2(c_8 +c_{9}+c_{10})+(s_w^4+c_w^4) c_{11}-2 c_w^2 s_w^2 c_{12}\nonumber\\
a_4^{ZZ}&=&(c_w^2-s_w^2)^2 c_{12}+4 c_w^2 s_w^2(c_{8}+ c_{9}+c_{10})-4 c_w^2 s_w^2 c_{11}\nonumber\\
a_5^{ZZ}&=&4 c_w^2 s_w^2(c_{13} +c_{14} +c_{15})+(s_w^4+c_w^4-2 c_w^2 s_w^2) c_{16}-4 c_w^2 s_w^2 c_{17}\nonumber\\
a_6^{ZZ}&=&(c_w^4+s_w^4) c_{17}+2 c_w^2 s_w^2(c_{13} +c_{14} +c_{15})-2 c_w^2 s_w^2 c_{16}.
\label{transform}
\eea

We have thus listed all operators that contribute to the $\gamma \gamma \gamma \gamma$ and  $\gamma \gamma ZZ$ vertices. Note that $c_1$ and $c_2$ do not appear in the RHS in eq.~\ref{transform} because the corresponding operators contribute only to the $ZZZZ$ coupling.  As we want to measure these couplings by the $\gamma \gamma \rightarrow \g\g$ and $\gamma \gamma \rightarrow ZZ$ processes respectively, let us  also list operators that might enhance non-local background contributions through the processes like $\gamma \gamma \to X^* \to \gamma \gamma/ZZ$ at the same order, where $X$ is some SM field. We find that the only  dimension 6 operators giving such non-local contributions at the same order are those that contribute via the $\gamma \gamma \to h^*\to ZZ$ processes due to the  anomalous $h \gamma \gamma$ couplings they introduce. These operators (already listed in Ref.~\cite{Buchmuller:1985jz}) are,
\beq
{\cal L}_6 = \frac{b_1}{\Lambda^2}|\Phi|^2 B_{\mu \nu}B^{\mu \nu}+ \frac{b_2}{\Lambda^2}|\Phi|^2 W^I_{\mu \nu}W^{I\mu \nu}+\frac{b_3}{\Lambda^2} (\Phi^\dagger \sigma_I \Phi) B_{\mu \nu}W^{I\mu \nu}.
\label{uno}
\eeq
which give the following dimension 5 operator once the Higgs field, $\Phi$, gets a VEV,
\bea
 \frac{a_0 v}{\Lambda^2}h F_{\mu \nu}F^{\mu \nu}
\eea
where $v=246$ GeV and,
\beq
a_0=c_w^2 b_1 +s_w^2 b_2-c_w s_w b_3.
\eeq
Note that heavy particles that interact with the Higgs boson and photons generally induce this operator when integrated out.  Fortunately, the coupling $a_0$ can be accurately measured at the LHC by measuring the $h\to \gamma \gamma$ partial width. Thus the effect of the only dimension 6 operator that contributes to the cross section can be subtracted.

We have not identified couplings for  $ \gamma \gamma X$  or  $ XZZ$, that arise from dimension $>6 $ operators and contribute to this process by an $X$ exchange, because these contributions would have the dependance $\sim 1/\Lambda^n$ with $n>4$ which would be of  higher order than dimension 8 which is the lowest order at which the QNGCs get a contribution. 

\subsection{Higgsless case}
\label{higgsless}

Note that our treatment in the previous subsection differs from that in Refs.~\cite{Dervan:1999as,Eboli:2000ad,Pierzchala:2008xc,Kepka:2008yx} where only the $\g\g\to ZZ$ process has been discussed. Among all the terms in eq.~\ref{Lph} the authors consider  only the operators ${\cal O}_1^{ZZ}$ and ${\cal O}_2^{ZZ}$ (i.e., the operators that have the coefficients $a_1^{ZZ}$ and $a_2^{ZZ}$ respectively). This can be justified if there is no light Higgs and electroweak symmetry breaking (EWSB) is non-linearly realized at low energies.   Let us see why this is so. We follow the construction of Burgess et al. (Ref.~\cite{Burgess:1992gx}), use  the matrix,
\bea
\Sigma = \exp(2i X_i \pi_i/v)
\eea
 and  the covariant derivative,
\beq
D_\mu \Sigma =\Sigma^\dagger D_\mu \Sigma -i \Sigma^\dagger [gW^a_\mu T_a+g' B_\mu Y] \Sigma. 
\eeq
to define  the following fields,
\bea
e {\cal A}_\mu &=&2i~ {\rm Tr}[X_{em}D_\mu \Sigma] \nonumber\\
\frac{g}{2c_w}{\cal Z}_\mu &=&iÊ~{\rm Tr}[X_3 D_\mu \Sigma] \nonumber\\
g {\cal W}^{\pm}_\mu &=&i \sqrt{2}~Ê{\rm Tr}[T_{\pm} D_\mu \Sigma]. 
\label{NL}
\eea
Here  $Y$ is the hypercharge generator, $T_{\pm}=T_1 \pm T_2$, where $T_i$ are the SU(2)$_L$ generators.  $X_{em}$ and $X_3$ are orthogonal linear combinations of $Y$ and $T_3$, $X_{em}$ being the unbroken generator of U(1)$_{em}$. We have kept the unconventional normalization of Ref.~Ê\cite{Burgess:1992gx}, \textit{viz} ${\rm Tr}[T_a, T_b]= \frac{1}{2}\delta_{ab}$, ${\rm Tr}[T_a, Y]= 0$ and ${\rm Tr}[Y^2]= \frac{1}{2}$. 

As shown in Ref.~Ê\cite{Burgess:1992gx},  the fields ${\cal A}$, ${\cal Z}$ and ${\cal W}^{\pm}$ in eq.~\ref{NL} transform purely electromagnetically and exactly like $A$, $Z$ and $W^{\pm}$ respectively. In the unitary gauge, $\Sigma \to 1$ so that  ${\cal A}\to A$, ${\cal Z}\to Z$ and ${\cal W}^{\pm}\to W ^{\pm}$. It is thus easy to construct gauge invariant  operators we are interested in if EWSB is non linearly realized. In the unitary gauge these are just all possible operators constructed from the $A$, $Z$ and $W^\pm$ fields that respect the U(1)$_{em}$ symmetry.    We get therefore for the $\g\g ZZ$ coupling,
\beq
{\cal L}^{Higgsless}_{QNGC} = \frac{(g/2c_w)^2a_1^{hl}}{\Lambda^2} F_{\mu \nu}F^{\mu \nu} Z_\rho Z^\rho+ \frac{(g/2c_w)^2 a_2^{hl}}{\Lambda^2}F_{\mu \nu}F^{\mu \rho} Z_\rho Z^\nu
\label{laghl}
\eeq 
Note that in this case we get as the lowest order contributions to QNGCs  two dimension 6 operators, which are same as as   ${\cal O}^{ZZ}_1$ and ${\cal O}^{ZZ}_2$ in eq.~\ref{Lph}, and none of the other operators in eq.~Ê\ref{Lph} are present above. Thus, unlike in the case with the  light Higgs boson,  these operators are indeed more important here, and this is why they are the only ones that appear in the analyses of Refs.~\cite{Dervan:1999as,Eboli:2000ad,Pierzchala:2008xc,Kepka:2008yx}. The $\g \g \g \g$ coupling does not get any contribution at this order.

Another way to understand the above fact is by using the goldstone boson equivalence theorem which states that at high energies longitudinal gauge boson production processes should have the same amplitude as processes in which the corresponding goldstone bosons are produced. The operators in eq.~\ref{laghl} arise from operators like $c~{\rm Tr}[(D_\mu \Sigma)^\dagger D^\mu \Sigma] F_{\mu \nu} F^{\mu \nu}$. This  can be expanded to give the terms involving the goldstones  like, $((c/\Lambda^2)(\partial_\rho \vec{\pi} \partial^\rho \vec{\pi}/v^2)) F_{\mu \nu}F^{\mu \nu}$, which tells us that the ${\cal A}(\g\g \to Z_L Z_L)$ amplitude will be ${\cal O}(c \hat{s}^2/(v^2 \Lambda^2))$, ignoring the dimensionless electroweak couplings.  This is larger than the amplitude due to an operator like $((c/{\Lambda^4}) Z_{\rho \sigma}Z^{\rho \sigma}) F_{\mu \nu}F^{\mu \nu}$, not included in eq.~Ê\ref{NL}, which will give ${\cal A}(\g\g \to Z Z)= {\cal O}(c \hat{s}^2/\Lambda^4)$. The crucial difference  is that, unlike  the light Higgs case, the goldstones  here are strongly coupled and suppressed by  factors of  $1/v$ and not $1/\Lambda$.

\subsection{Graviton exchange in extra-dimensional theories as a source of QNGCs}
\label{extrad}

In extra-dimensional theories where the fundamental gravity scale can be a few TeV, the graviton is accompanied by Kaluza-Klein (KK) partners in the 4D effective theory. Exchange of the $(4+\delta)$-dimensional graviton, $\delta$ being the number of extra dimensions, can be thought of as the excahnge of the 4D graviton and its massive KK partners. The effective operator induced by tree-level graviton exchange is given by~\cite{Giudice:1998ck},
\beq
{\cal O}_T =\frac{4 \pi}{\Lambda_T^4}\left(\frac{T_{\mu \nu}T^{\mu \nu}}{2}-\frac{1}{\delta + 2}\frac{T^\mu_\mu T^\nu_\nu}{2}\right)
\label{vg}
\eeq
where $T^{\mu \nu}$ is the energy-momentum tensor. At tree level only dimension 8 operators are induced (at loop level  only one dimension 6 operator operator is induced by virtual graviton exchange but this is a four fermion operator not involving the gauge bosons or the Higgs~\cite{Giudice:2003tu}). 

Almost all the operators in eq.~\ref{L8} can be obtained by expanding $T^{\mu \nu}$ in eq.~\ref{vg}. To show this let us write down the energy-momentum tensor for the $B_\mu$ and $W^I_\mu$ gauge bosons and the Higgs boson,
\bea
T^{\mu \nu}_B&=& -B^{\mu\rho} B^\nu_\rho+\frac{1}{4}g^{\mu\nu} B^{\rho \sigma}B_{\rho\sigma}
\nonumber\\
T^{\mu \nu}_W&=& -W^{I\mu\rho} W^{I\nu}_\rho+\frac{1}{4}g^{\mu\nu} W^{I\rho  \sigma}W^I_{\rho\sigma}
\nonumber\\
T^{\mu \nu}_\Phi&=&D^{\mu} \Phi^\dagger D^{\nu}\Phi+D^{\nu} \Phi^\dagger D^{\mu}\Phi-g^{\mu\nu}(D^{\mu} \Phi^\dagger D_{\mu}\Phi- m^2 \Phi^\dagger \Phi).
\label{em}
\eea

Note that virtual graviton exchange will also generate operators involving the gluon field strength, $G^I_{\mu \nu}$, like $G^I_{\mu \nu}G^{I\mu \nu} B_{\rho \sigma}B^{\rho \sigma}$, $G^I_{\mu \nu}G^{I\mu \nu}D^{\rho} \Phi^\dagger D_{\rho}\Phi$ etc. Such operators would enhance the signal by contributing to the central exclusive pomeron fusion   process (CEP), $pp(CEP\to \g\g/ZZ)pp$.  The luminosity of photons produced by the protons is however higher than the luminosity of the pomerons produced that undergo exclusive fusion (by exclusive we mean that the pomerons do not disintegrate into fragments) by a few orders of magnitude at the high energies where these operators become important (see Fig. 2 in Ref.~\cite{Khoze:2001xm}). Thus the $pp(CEP\to \g\g/ZZ)pp$ contribution is expected to be negligible compared to the $pp(\g\g\to \g\g/ZZ)pp$ contribution. In any case any contribution form this channel would only enhance the signal and thus improve the experimental potential of observing effects of  virtual graviton exchange.

\section {Constraints}

QNGCs are very weakly constrained by existing data. There are no constraints on $\g\g\g\g$ couplings and the only constraints are on $\g\g ZZ$ couplings. We  first consider the light Higgs case discussed in Section~Ê\ref{lh}. A LEP analysis~\cite{Abbiendi:2004bf} based on the $e^+ e^-\to Z\gamma\gamma$ process puts the following constraints on the operators ${\cal O}_1$ and ${\cal O}_2$ in eq.~\ref{Lph},  
\beq
-\frac{1}{(69\, {\rm GeV})^4} <\frac{a^{ZZ}_1 }{\Lambda^4}<\frac{1}{(93\, {\rm GeV})^4}
\eeq
and,
\beq
-\frac{1}{(65\, {\rm GeV})^4}<\frac{a^{ZZ}_2 }{\Lambda^4}<\frac{1}{(65\, {\rm GeV})^4}.
\eeq
While the authors of Ref.~\cite{Abbiendi:2004bf} did not carry out their analysis for the other  operators in eq.~\ref{Lph}, as these are also dimension 8 operators we expect their contribution to these processes to be of a similar magnitude. Thus the constraints on these couplings are also expected to be very weak. Somewhat stronger constraints can be derived from electroweak precision data. In Ref.~\cite{Eboli:2000ad} precision constraints on the operators are derived and they find  the bounds, $|a^{ZZ}_{1,2}/\Lambda^4|\lesssim 1/(270$ GeV)$^{-4}$, which, as we shall see later, are still far too weak compared to the expected LHC sensitivity.  

As the operators in the higgsless case discussed in Section~Ê\ref{higgsless} are exactly the two operators discussed above, the only difference being that we use a different parametrization for the couplings, the same constraints can be translated to the couplings in eq.~\ref{laghl} in the  higgsless case,
\beq
-\frac{1}{(27\, {\rm GeV})^2} <\frac{a^{hl}_1 }{\Lambda^2}<\frac{1}{(50\, {\rm GeV})^2}
\eeq
and,
\beq
-\frac{1}{(24\, {\rm GeV})^2}<\frac{a^{hl}_2 }{\Lambda^2}<\frac{1}{(24\, {\rm GeV})^2}, 
\eeq
whereas the precision constraints in Ref.~\cite{Eboli:2000ad}  imply $|a^{hl}_{1,2}/\Lambda^2|\lesssim 1/(420$ GeV)$^{-2}$. 

Now we discuss the constraints on the scale, $\Lambda_T$, for virtual graviton exchange, which appears  in eq.~Ê\ref{vg}. The   strongest constraints on $\Lambda_T$ come from LHC data at 7 TeV. With 36 pb$^{-1}$ CMS data at 7 TeV the $pp \to jj$ process can be used to derive the constraint $\Lambda_T>3.8$ TeV~\cite{Franceschini:2011wr} at 95 $\%$ confidence level. The same process puts the constraint $\Lambda_T>3.6$ TeV~\cite{Franceschini:2011wr} with 36 pb$^{-1}$ ATLAS data  at 95 $\%$ confidence level. With 1.1 fb$^{-1}$ CMS data, the $pp \to \g\g$ process puts the  weaker constraint $\Lambda_T>3.1$ TeV at  95 $\%$ confidence level~\cite{exdpas},  but this process is eventually expected to probe scales up to about $\Lambda_T=6$ TeV~\cite{Giudice:1998ck}.

Finally,  consider the operators in eq.~\ref{uno}. At tree level the couplings $b_1$ and $b_2$ in eq.~\ref{uno} renormalize the coefficient of the kinetic terms for the gauge bosons $B_\mu$ and $W^I_\mu$ which is equivalent to a renormalization of the couplings $g'$ and $g$. Thus all tree level effects due to the $b_1$ and $b_2$ can be absorbed in a redefinition of the couplings and hence these couplings are unconstrained. The coupling $b_3$ is related to the $S$-parameter by~\cite{Han:2004az},
\beq
\frac{b_3}{\Lambda^2}=\frac{\alpha_{em}}{4 s_w c_w v^2} \Delta S 
\eeq
Here $\alpha_{em}$ is the fine structure constant $v=246$ GeV and $s_w, c_w$ are the sine and cosine of the weak mixing angle. The bound on the $S$-parameter for $m_h=113$ GeV and with no restrictions on the $T$ parameter is $|\Delta S| \lesssim 0.3$~\cite{Han:2004az} at 90$\%$ confidence level. This translates to the following bound on $b_3$,
\beq
\left|\frac{b_3}{\Lambda^2}\right|<\frac{1}{(6.6\, {\rm TeV})^2}.
\eeq
This coupling is also constrained by measurements of the triple gauge couplings but these constraints are far weaker~\cite{Dutta:2007st}.
 \section{High energy behavior of amplitudes and violation of unitarity at tree-level}
\label{unitarity}

First let us look at the $\gamma \gamma \to \gamma \gamma$ process. We can find out the high energy behavior by dimensional analysis. The high energy behavior of  the contribution from the local operators in eq.~\ref{Lph} differs from the contribution from the non-local process, where the $h\gamma\gamma$ vertex is derived from the operator ${\cal O}_0$, as follows,
\bea
{\cal O}^{\g\g}_1,{\cal O}^{\g\g}_2 &:&{\cal A}(\gamma \gamma \to \gamma \gamma)\sim  a_i \frac{\hat{s}^2}{\Lambda^4}
\nonumber\\
{\cal O}_0 &:&{\cal A}(\gamma \gamma \to h^*\to \gamma \gamma)\sim a_0 \frac{v\hat{s}}{\Lambda^2}\frac{1}{\hat{s}}\times a_0 \frac{v\hat{s}}{\Lambda^2} \sim a_0^2 \frac{v^2\hat{s} }{\Lambda^4}
\nonumber\\
\eea
where $\hat{s}$ is the photon-photon center of mass energy squared. The local contribution is thus expected to dominate over the non-local contribution at high energies.

For the $\gamma \gamma \to ZZ$ process  the operators in the light Higgs case in  eq.~\ref{Lph}  can be divided into into two categories according to the final polarization of the $Z$s. At high energies the operators ${\cal O}^{ZZ}_1,~{\cal O}^{ZZ}_2$ contribute mainly to the production of longitudinally polarized $Z$-bosons through the process $\gamma \gamma \rightarrow Z_L Z_L$ while  the operators ${\cal O}^{ZZ}_3-{\cal O}^{ZZ}_6$ contribute mainly to transverse $Z$ production through the process $\gamma \gamma \rightarrow Z_T Z_T$. This can be understood by using the goldstone boson equivalence theorem. The operators like ${\cal O}^{ZZ}_3-{\cal O}^{ZZ}_6$ do not arise from dimension 8 operators involving the Higgs field (see eq.~\ref{transform}) and so they do not introduce new couplings to the Goldstone bosons (that are eaten by the gauge bosons in the unitary gauge). New contributions to the process $\gamma \gamma \rightarrow Z_L Z_L$ they introduce are, therefore,  suppressed.

Let us now see the energy dependance of the $\gamma \gamma \rightarrow Z Z$ amplitude of the dominant $Z$-polarization modes for the different operators using dimensional analysis in the high energy limit,
\bea
{\cal O}^{ZZ}_3-{\cal O}^{ZZ}_6 &:&{\cal A}(\gamma \gamma \to Z_T Z_T)\sim  a_i \frac{\hat{s}^2}{\Lambda^4}
\nonumber\\
{\cal O}^{ZZ}_1,{\cal O}^{ZZ}_2 &:& {\cal A}(\gamma \gamma \to Z_L Z_L)\sim a_i \frac{M_Z^2 \hat{s}}{\Lambda^4}\frac{\hat{s}}{M_Z^2} \sim a_i \frac{\hat{s}^2}{\Lambda^4}
\nonumber\\
{\cal O}_0 &:&{\cal A}(\gamma \gamma \to h^*\to Z_T Z_T)\sim a_0 \frac{v\hat{s}}{\Lambda^2}\frac{1}{\hat{s}}\frac{g M_Z}{c_w}\sim a_0 \frac{M_Z^2}{\Lambda^2}
\nonumber\\
{\cal O}_0 &:&{\cal A}(\gamma \gamma \to h^*\to Z_L Z_L)\sim a_0 \frac{v\hat{s}}{\Lambda^2}\frac{1}{\hat{s}}\frac{g M_Z}{c_w}\frac{\hat{s}}{M_Z^2}\sim a_0 \frac{\hat{s}}{\Lambda^2}.
\nonumber\\
\label{hen}
\eea
where the the $\hat{s}/M_Z^2$ factor for the longitudinal modes comes from the longitudinal polarization vectors and $g M_Z/c_w$ is the SM $hZZ$ coupling. Note that according to eq.~\ref{hen} the $\gamma \gamma \to h^* \to ZZ$ process would mainly produce longitudinal $Z$s. As discussed earlier the contribution of the operator ${\cal O}_0$  can be subtracted by  measuring the $h\to \gamma \gamma$ partial decay width. For the operators in the higgsless case in Section~Ê\ref{higgsless},  the dominant mode will  be  $\gamma \gamma \rightarrow Z_L Z_L$ and the energy dependence would be, 
\beq
{\cal A}(\gamma \gamma \to Z_L Z_L)\sim (g/2c_w)^2 a^{hl}_i \frac{ \hat{s}}{\Lambda^2}\frac{\hat{s}}{M_Z^2}\sim  a^{hl}_i \frac{\hat{s}^2}{\Lambda^2 v^2}.
\eeq

As all the amplitudes above  grow with energy they would all violate partial wave unitarity for some value of $\hat{s}$. We obtain the perturbative unitarity bound for the processes in Appendix B. The  condition that perturbative unitarity is not violated is,
\bea
( {\rm Re}(b_l))^2+\beta \sum_{\epsilon_3,\epsilon_4} |a_l|^2+\delta_l<\frac{1}{4},
\label{UB}
\eea
where $a_l~(b_l)$ is the $l$-th partial wave amplitude for the $\g\g \to ZZ(\g\g)$ process, $\beta=\sqrt{1-\frac{4 M_Z^2}{\hat{s}^2}}$, $\hat{s}$ is the photon-photon center of mass energy, $\delta_l$ is the positive contribution from other processes and $\epsilon_3$ and $\epsilon_4$ are the polarizations of the $Z$ bosons produced.  For the first term the final polarizations are same as the initial. The initial polarizations of the photons must be chosen to maximize the LHS to get the most stringent possible bound. We find the  most stringent bounds from the $l=0$ mode. 

\begin{figure}[t]
\begin{center}
\includegraphics[width=0.9\columnwidth]{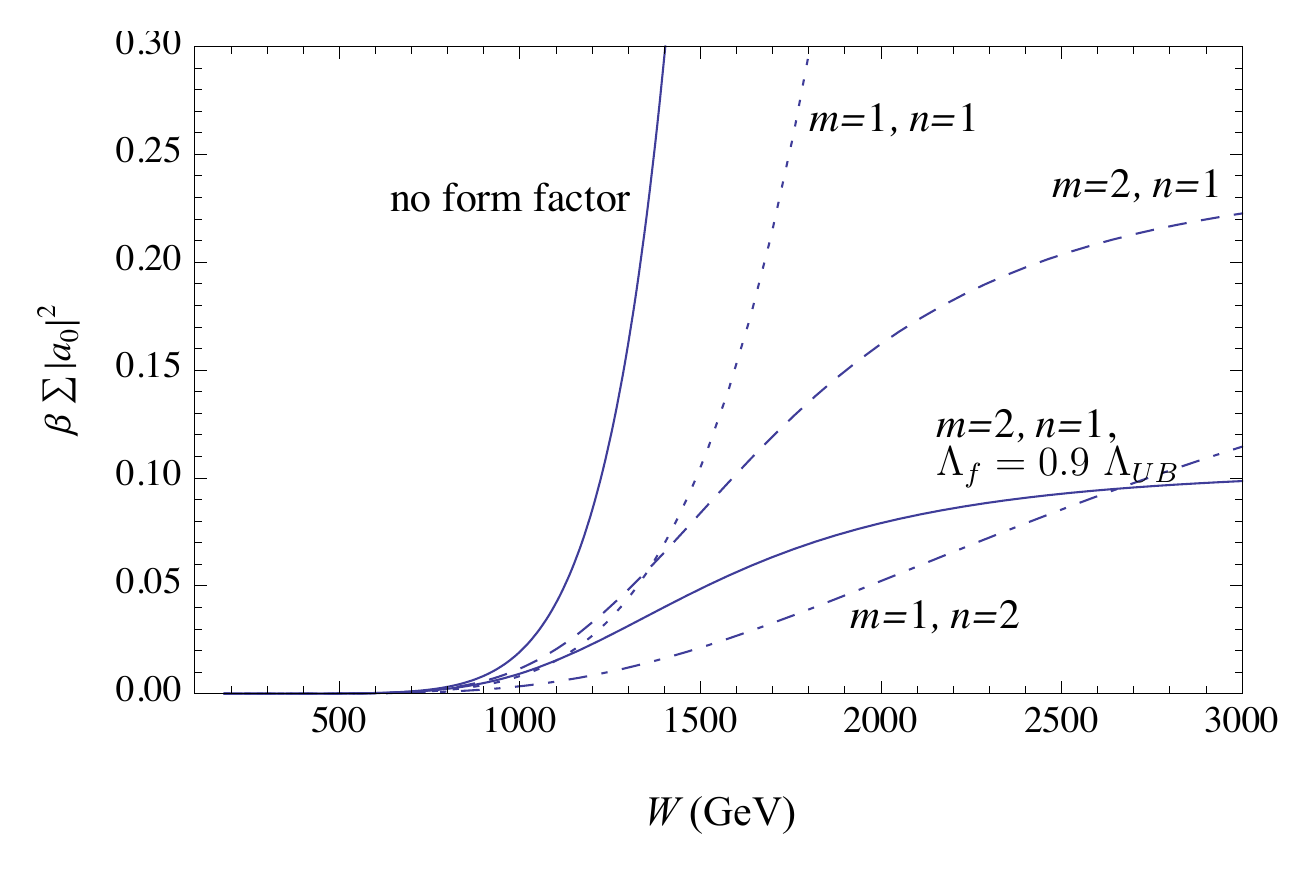}
\end{center}
\caption{Here we plot $\left( ( {\rm Re}(b_l))^2+\beta \sum_{\epsilon_3,\epsilon_4} |a_l|^2 \right)$ in eq.~\ref{UB} vs the photon-photon center of mass energy for  $l=0$  with and without the form factor in eq.~\ref{ff} taking $b_i/\Lambda^2=1$ TeV$^{-2}$ and $c_i/\Lambda^4=1$ TeV$^{-4}$. We take $\Lambda^{\g\g/ZZ}_f=\Lambda^{\g\g/ZZ}_{UB}$ in the form factor for all the different cases  other than the lower solid line where we take $\Lambda^{\g\g/ZZ}_f=0.9~ \Lambda^{\g\g/ZZ}_{UB}$.}
\label{form}
\end{figure} 

To ameliorate the growth of the amplitude with energy we can use  form factors as follows,
\beq
{\cal A}\rightarrow {\cal A}\left(\frac{ 1}{1+(\hat{s}/\Lambda_f^2)^m}\right)^{n}.
\label{ff}
\eeq
By Taylor expanding the modified amplitude we see that by introducing a form factor we effectively introduce higher order contributions, such as those expected from loop effects and higher dimensional operators, to cancel  the tree-level growth of the amplitude. For example if ${\cal A}=k \hat{s}^2/\Lambda^4$ the for the choice $m=2$ and $n=1$ the modified amplitude is , 
\beq
\frac{k\hat{s}^2/\Lambda^4}{1+(\hat{s}/\Lambda_f^2)^2}=(k\hat{s}^2/\Lambda^4)\left(1-(\hat{s}/\Lambda_f^2)^2+(\hat{s}/\Lambda_f^2)^4-(\hat{s}/\Lambda_f^2)^6...\right).
\eeq
In order that the eq.~\ref{UB} is obeyed we would require that,
\bea
( {\rm Re}(b_l))^2<0.1\label{UB2}\\
\beta \sum_{\epsilon_3,\epsilon_4} |a_l|^2<0.1.
\label{UB3}
\eea
The RHS in the two equations above do not add up to the RHS of eq.~\ref{UB} because we have made some allowance for other contributions  to $\delta_l$. To ensure  that these conditions are obeyed we use form factors for both the $\g\g \to\g\g$ and the $\g\g \to ZZ$ amplitudes.  

Fig.~\ref{form} shows the growth of $\left( ( {\rm Re}(b_l))^2+\beta \sum_{\epsilon_3,\epsilon_4} |a_l|^2 \right)$ in eq.~\ref{UB}  for  $l=0$  with energy, for different choices of the form factor parameters. We consider the light Higgs case in Section~Ê\ref{lh}, taking the couplings $b_i/\Lambda^2=1$ TeV$^{-2}$ and $c_i/\Lambda^4=1$ TeV$^{-4}$. Let $\Lambda^{\g\g}_{UB}$ and $\Lambda^{ZZ}_{UB}$ be the values of $\hat{s}$ where  the conditions in eq.~\ref{UB2} and~\ref{UB3} are respectively violated when no form factor is applied.  We can see that the amplitude keeps growing for $m=n=1$ and  $\Lambda^{\g\g/ZZ}_f=\Lambda^{\g\g/ZZ}_{UB}$,  thus violating the perturbative unitarity bound.  However, the amplitude is suppressed below the bound for $m=1,n=2$ and $m=2,n=1$ for the same values of $\Lambda^{\g\g/ZZ}_f$. We see that in the latter case the amplitude saturates the bounds in eq.~\ref{UB2} and~\ref{UB3} at high energies. We also show a curve with $m=2,n=1$ but $\Lambda_f^{\g\g/ZZ}=0.9~ \Lambda^{\g\g/ZZ}_{UB}$ which coincides with the $\Lambda_f^{\g\g/ZZ}= \Lambda_{UB}^{\g\g/ZZ}$ curve at low energies  but deviates from it  for $\hat{s}$ close to $\Lambda^2_{UB}$.  
Unless otherwise mentioned from now on we will use  form factors with $m=2,n=1$ and $\Lambda^{\g\g/ZZ}_f=\Lambda^{\g\g/ZZ}_{UB}$. While our final results will depend on this specific choice of form factor a different form factor would result in a cross-section with a different numerical value but the same order of magnitude.  Thus there will be a relatively small difference in our final sensitivity results on $\Lambda$ as the cross-section goes as $\sigma \sim \Lambda^{-8}$ ($\sigma \sim \Lambda^{-4}$) for the dimension 8 (dimension 6) operators in the light Higgs (higgsless) case.

\begin{figure}[t]
\begin{center}
\includegraphics[width=0.9\columnwidth]{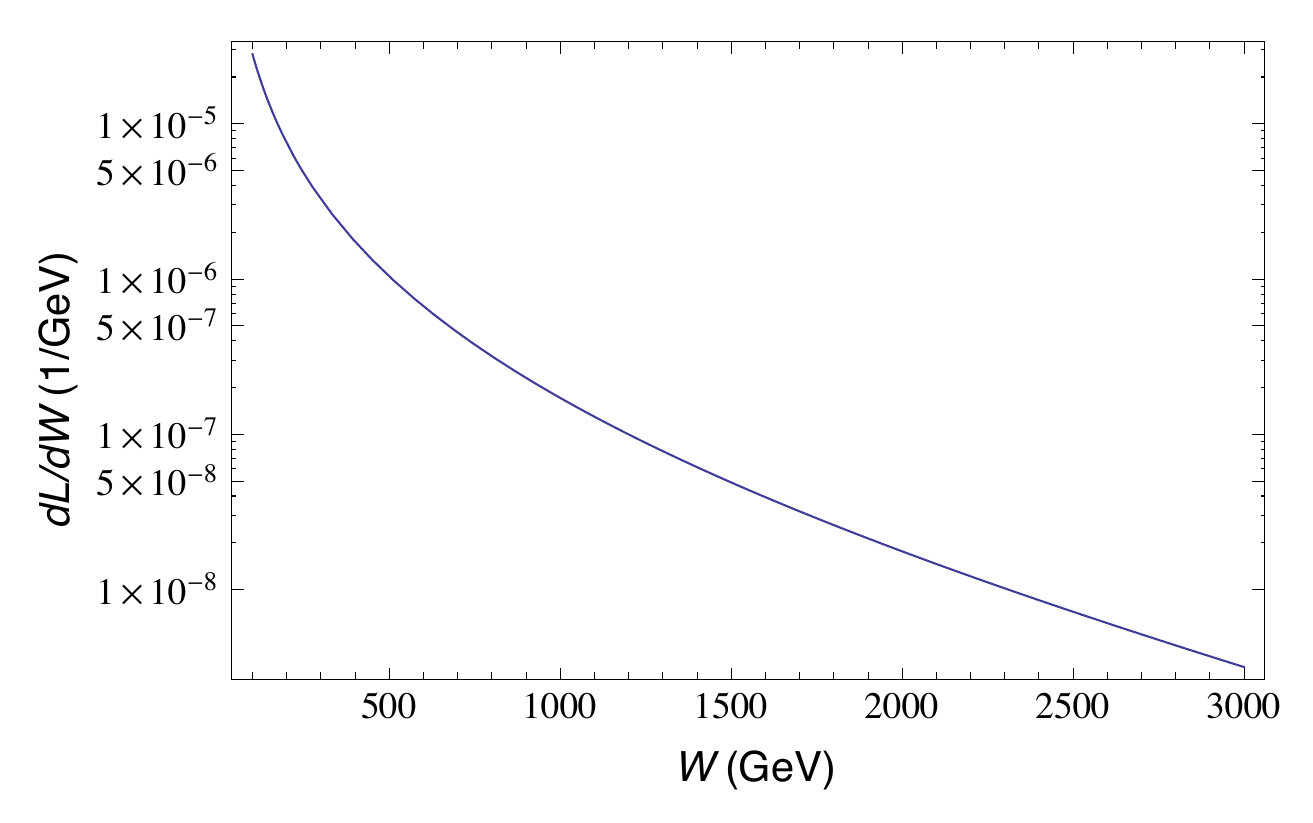}
\end{center}
\caption{The Luminosity function $dL/dW$ in eq.~\ref{lf} taking $q^2_{max}=2$ GeV$^2$.}
\label{lumin}
\end{figure}

 \section{The Equivalent Photon Approximation and the proton level cross-section for $pp(\gamma \gamma \rightarrow Z Z)pp$}

\label{epha}

Consider a general process $pp(\gamma\gamma \to X)pp$. To find the amplitude for this process we need to know the $pp\gamma$ vertex. From gauge invariance the most general form for this vertex is~\cite{Close:2007zzd},
\beq
-i e \left[F_1 (q^2)\gamma^\mu+\frac{\mu_p-1}{2 m_p} F_2(q^2) i \sigma^{\mu \nu}q_\nu\right].
\eeq
Here $e\mu_p/2 m_p$ is the proton magnetic moment with $\mu_p^2=7.78$, $q_i^2$,  the invariant mass of one of the  photons is, as shown in Appendix C,  always  space-like  in such a process and thus it is negative. The functions $F_1(q^2)$ and $F_2(q^2)$  in the vertex can be expressed in terms of the the empirically determined elastic electric and magnetic form factors for the proton, $G_E(q^2)$ and $G_M(q^2)$ respectively, as follows,
\bea
F_1(q^2)&=&\frac{G_E + \tau G_M}{1 + \tau}~~~~~F_2(q^2)=\frac{G_M-G_E}{\kappa(1+\tau)}
\nonumber\\
G_E&=&G_M/\mu_p=(1-q^2/q_0^2)^{-2}.
\label{emform}
\eea
Here $G_E (q^2)$ and $G_M (q^2)$ have been written in the dipole approximation with  $q_0^2=0.71 ~{\rm GeV}^2$ and $\tau=(-q^2)/4 m_p^2$. By a change of variables the final phase space integration for the process $pp(\gamma\gamma \to X)pp$ can be done over $d(-q^2_1) d(-q^2_2) d\omega_1d\omega_2$,  instead of the usual variables~\cite{Budnev:1974de}, $\omega_i$ being the energy of the photons. The cross section thus obtained would receive most of the contribution from the region in phase space where the $|q_i^2|$ are small (this also corresponds to small scattreing angles for the proton) because of the $1/q_i^2$ factors from the photon propagator. Note that, there is a   kinematic lower  bound on $|q_i^2|$,
\beq
 q_i^2 \lesssim - \frac{m^2 \omega_i^2} {E(E-\omega_i)}
\label{kin}
\eeq
where $E$ is the energy of the proton in the center of mass frame and $m_p$ its mass (see Appendix C for the derivation). 

The fact that most of the contribution to the cross section comes from the small $|q_i^2|$ region means that we can evaluate  the contribution to the amplitude from the $\gamma \gamma \to X$ part of the diagram in the $|q_i^2|\to 0$ limit. This is the so-called  Equivalent Photon Approximation (EPA). This amounts to treating the photons as real with only transverse polarizations while doing the the $\gamma \gamma \to X$ part of the calculation so that the total cross section can be written in  the factorized form,
\beq
 \sigma = S^2_{QED} \int_{2 M_Z}^{{W}_{max}} \frac{dL}{dW}\sigma_{\gamma\gamma} d{W}.
\label{epa}
\eeq
Here $W=\sqrt{\hat{s}}$ is the photon-photon center of mass energy and $\sigma_{\gamma\gamma}$ is the photon level cross section. $S^2_{QED}$, the survival probability for diffractive photon exchange processes, is the probability that the proton remains intact and is not broken due to subsequent inelastic QCD interactions. We take $S^2_{QED}=0.9$ following the theoretical calculation in Ref.~\cite{Khoze:2001xm}.  The function $dL/d{W}$ contains all the details of the proton electromagnetic form factors and also the integral over $1/q_i^2$ factors of the photon propagators.
A detailed calculation of $dL/d{W}$ using EPA leads to the following expressions (see Appendix D in Ref.~\cite{Budnev:1974de}),
\bea
\frac{dL}{d{W}}&=&\int_0^1 2 {W} f(x)f\left(\frac{W^2}{xs}\right)\frac{dx}{xs}\nonumber\\
f(x)&=&\frac{\alpha}{\pi}\frac{E}{\omega}\int_{q^2_{min}}^{q^2_{max}}\frac{d (-q^2)}{|q^2|}\left[\left(1-\frac{\omega}{E}\right)\left(1-\left|\frac{q^2_{min}}{q^2}\right|\right)D+\frac{\omega^2}{2 E^2}C\right]\nonumber\\
C&=&G_M^2~~~~~D=(4 m_p^2 G_E^2- q^2G_M^2)/(4 m_p^2- q^2)
\label{lf}
\eea
Here $x=\frac{\omega}{E}$ and $s=4E^2$. 
While the lower limit of the integration is set by kinematics (see eq.~\ref{kin}) we take the upper limit to be $q^2_{max}=2$ GeV. Beyond  $q^2_{max}=2$ GeV, the form factors in eq.~\ref{emform} become very small so that the contribution to the integral is negligible. 

To understand the physical meaning of $dL/d{W}$ we can multiply both sides of eq.~\ref{epa} by ${\cal L}_p$, the proton luminosity. Then we find that the luminosity function is the ratio of the differential photon luminosity $d{\cal L}_\gamma/d{W}$ and the proton luminosity,
\beq
dL/d{W}=\frac{d{\cal L}_\gamma/d{W}}{{\cal L}_p}.
\eeq
Note that here  $L$ is unitless and ${\cal L}_{\g,p}$ has the usual units $m^{-2}s^{-1}$. We plot the photon luminosity function in Fig.~\ref{lumin}. We find that,
\beq
S^2_{QED} \int_{2M_Z}^{2E}  \frac{dL}{d{W}} d{W} \sim 1.3\times10^{-3}.
\label{est}
\eeq
For a particular process $\g\g \to X$ this number gives  an upper bound on the ratio,
\beq
\frac{\sigma_{\gamma \gamma}\left(pp(\g\g \to X)pp\right)}{\sigma (\g\g \to X)}
\nonumber
\eeq
 if $\sigma_{\gamma \gamma}$  is a constant or  decreasing function of $W$ as is the case usually  for SM processes. Thus from a knowledge of $\sigma_{\gamma \gamma}(\g\g \to X)$ one can estimate $\sigma \left(pp(\g\g \to X)pp\right)$ using eq.~\ref{est}

 \section {Theoretical cross sections}
 In this section we present   the cross section for $pp(\gamma \gamma \rightarrow \g\g)pp$ and $pp(\gamma \gamma \rightarrow ZZ)pp$ (see Fig.~Ê\ref{fd}) with the proton-proton center of mass energy equal to 14 TeV. We will consider only the light higgs case in Section~Ê\ref{lh} taking all $b_i/\Lambda^2=1$ TeV$^{-2}$ and all $c_i/\Lambda^4=1$ TeV$^{-4}$ in eqs.~\ref{uno} and~\ref{L8}. With these values for the couplings and using eq.~\ref{UB2} and ~\ref{UB3} for $l=0$, we get the unitarity bound
$\Lambda^{\g\g}_{UB}=1220 {\rm ~GeV}$ and $\Lambda^{ZZ}_{UB}=1260 {\rm ~GeV}$ respectively.  

 \label{th}
 \begin{figure}[t]
\begin{center}
\includegraphics[width=0.8\columnwidth]{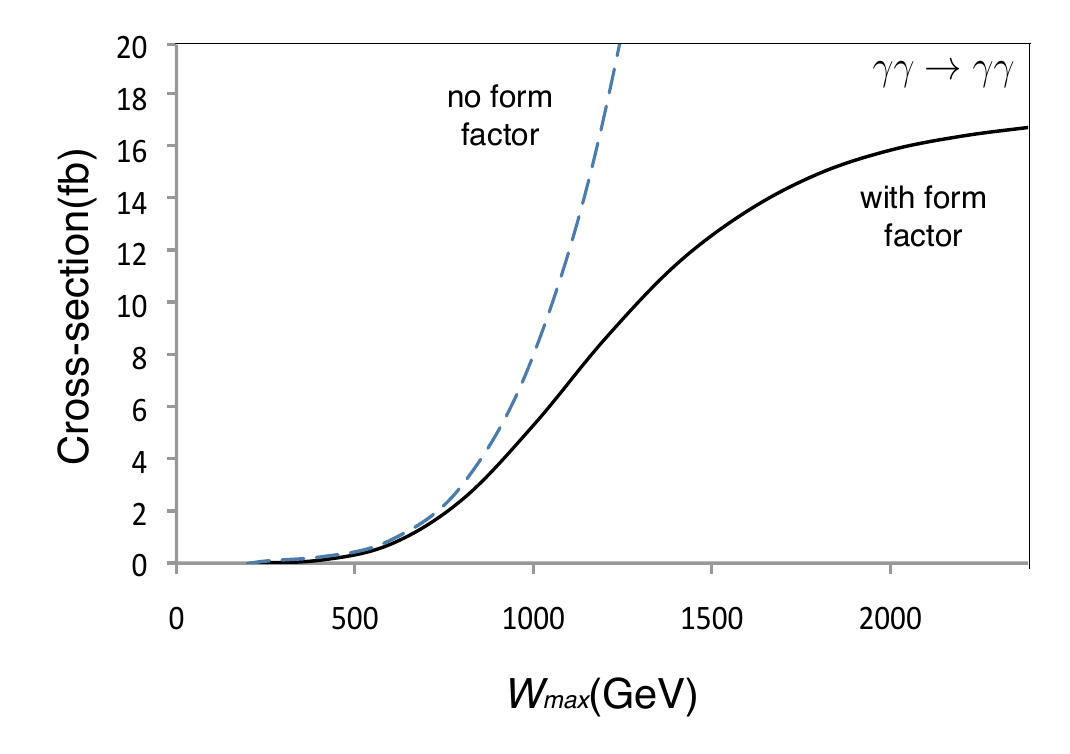}
\end{center}
\caption{The $pp(\gamma \gamma \rightarrow \g\g)pp$ cross section we obtain as a function of $W_{max}$ with and without a form factor. For the form factor we use in eq.~\ref{ff} with $m=2,n=1$ and   $\Lambda^{\g\g/ZZ}_f=\Lambda^{\g\g/ZZ}_{UB}$.   We have taken $b_i/\Lambda^2=1$ TeV$^{-2}$ and $c_i/\Lambda^4=1$ TeV$^{-4}$in eq.~\ref{transform}, the Higgs mass $m_h =120$ GeV and the proton-proton center of mass energy equal to 14 TeV.  }
\label{aa}
\end{figure}

\begin{figure}[t]
\begin{center}
\includegraphics[width=0.8\columnwidth]{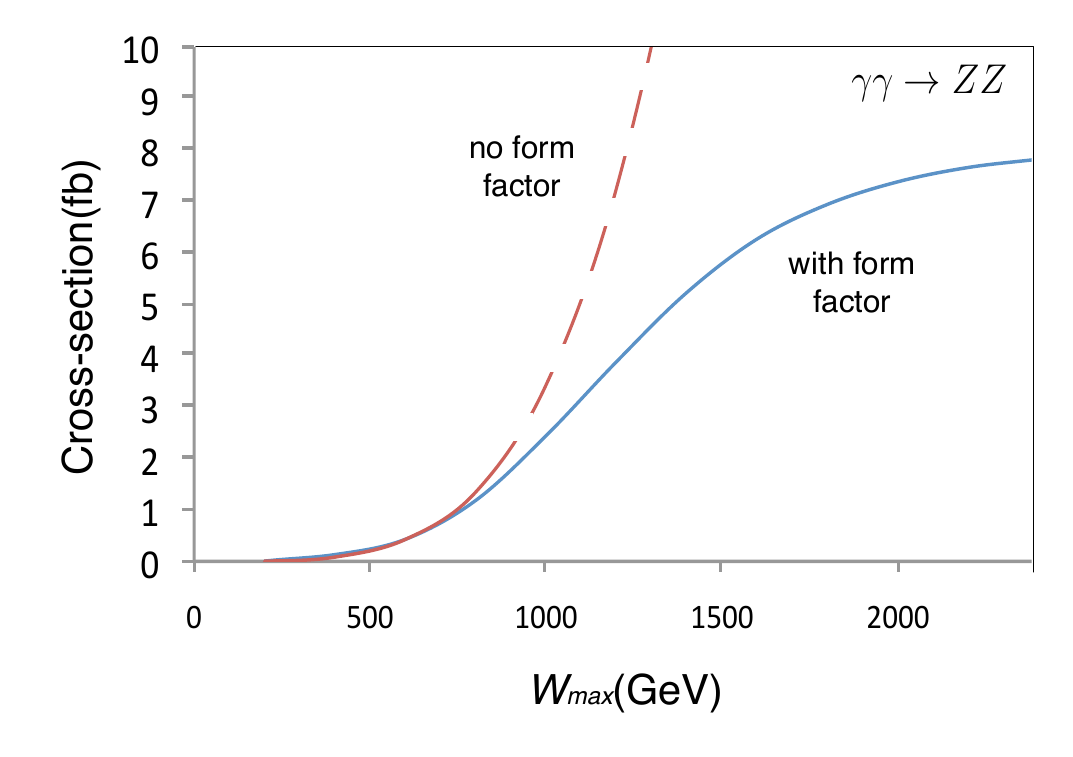}
\end{center}
\caption{The $pp(\gamma \gamma \rightarrow Z Z)pp$ cross section we obtain as a function of $W_{max}$ with and without a form factor. For the the form factor  we use in eq.~\ref{ff} with $m=2,n=1$ and   $\Lambda^{\g\g/ZZ}_f=\Lambda^{\g\g/ZZ}_{UB}$.  We have taken $b_i/\Lambda^2=1$ TeV$^{-2}$ and $c_i/\Lambda^4=1$ TeV$^{-4}$ in eq.~\ref{transform}, the Higgs mass $m_h =120$ GeV and the proton-proton center of mass energy equal to 14 TeV. }
\label{FF}
\end{figure}
\begin{figure}[t]
\centering
\hspace{-0.4 in}
\begin{tabular}{cc}
\includegraphics[width=0.5\columnwidth]{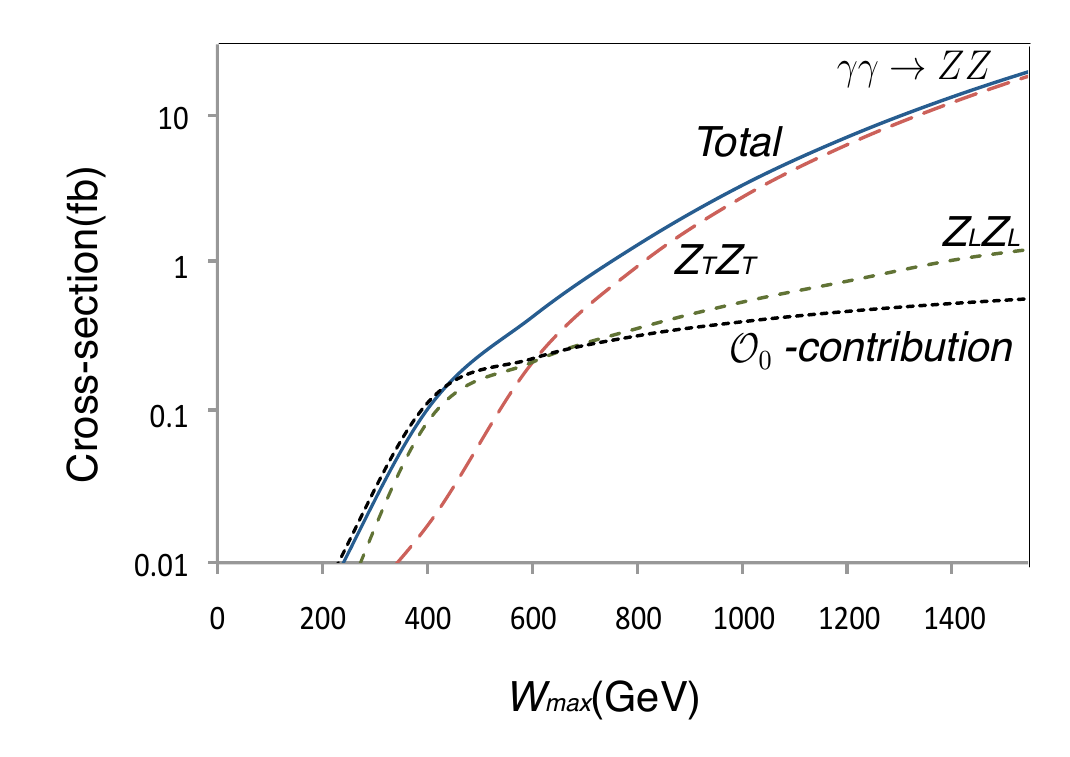} &
\includegraphics[width=0.5\columnwidth]{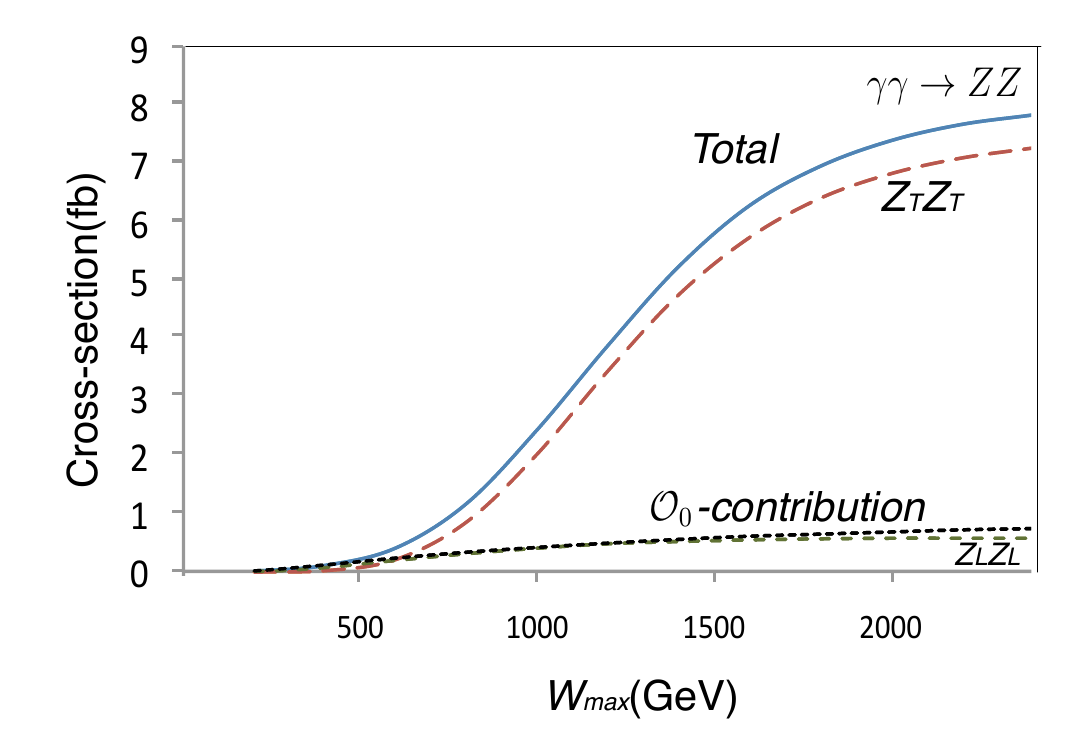} 
\end{tabular}
\caption{The $pp(\gamma \gamma \rightarrow Z Z)pp$ cross section we obtain as a function of $W_{max}$ without any form factor.   We have taken $b_i/\Lambda^2=1$ TeV$^{-2}$ and $c_i/\Lambda^4=1$ TeV$^{-4}$in eq.~\ref{transform}, the Higgs mass $m_h =120$ GeV and the proton-proton center of mass energy equal to 14 TeV.  We show the total cross section (solid), the $Z_T Z_T$ production cross section (big dashes), the $Z_L Z_L$ production cross section (small dashes) and the total cross section only due to the operator ${\cal O} _0$ (dotted) through the $\gamma\gamma\to h^*\to ZZ$ process. }

\label{bars}
\end{figure}

We have evaluated the cross section  with and without the form factor in eq.~\ref{ff}. For the form factor we have taken $\Lambda_f^{\g\g/ZZ}=\Lambda_{UB}^{\g\g/ZZ}$, $m=2$ and $n=1$. We have taken the Higgs mass $m_h =120$ GeV.  We have evaluated the cross section with the cut ${W}<W_{max}$ and varied ${W}_{max}$. This cut is important as the ambiguities due to the choice of form factor become more important for large values of $W_{max}$. For the  $pp(\gamma \gamma \rightarrow \g\g)pp$ process we show the results with and without the form factor in Fig.~\ref{aa}.  We have checked that the contribution from the $\g\g\to h^* \to\g\g$ process due to the presence of the operator ${\cal O}_0$  is small compared to the total cross-section as is expected from the arguments in Section~\ref{unitarity}. 

We show the results for the  $pp(\gamma \gamma \rightarrow ZZ)pp$ process with and without the form factor in Fig.~\ref{FF}.  In Fig.~\ref{bars}(left) we  show the  $\gamma \gamma \rightarrow ZZ$ cross section we obtain as a function of ${W}_{max}$ without any form factor. We show separately in the same figure the contribution due to  the operator ${\cal O}_0$ through the process $\gamma\gamma \to h^*\to ZZ$. We see that the non-local contribution due to  ${\cal O}_0$ dominates at low energies. The contribution to the cross section due to the other operators, however, grows more rapidly with $\hat{s}$ (as $\sigma_{\gamma\gamma}\sim \hat{s}^3$) compared to the ${\cal O}_0$ contribution. At higher energies ${\cal O}^{ZZ}_3$ contributes most to the cross-section. As mentioned earlier the coupling $a_0$ can be accurately measured by measuring the $h\to \gamma\gamma$ partial width so any deviation would indicate the presence of higher dimensional operators. We also show in  Fig.~\ref{bars}(left), the $\gamma \gamma \rightarrow Z_LZ_L$ and $\gamma \gamma \rightarrow Z_TZ_T$ contributions to the cross section. As explained before, for longitudinal $Z$ production, the main contribution comes from the operators ${\cal O}_0, {\cal O}^{ZZ}_{1,2}$ with ${\cal O}_0$ contributing dominantly at low energies and ${\cal O}^{ZZ}_1$  contributing dominantly at higher energies. For transverse $Z$ production only the operators ${\cal O}_0$, ${\cal O}^{ZZ}_{3,4,5,6}$ contribute significantly with the dominant contribution coming from ${\cal O}^{ZZ}_3$. 

In Fig.~\ref{bars}(right) we have the signal cross section curves as in Fig.~\ref{bars}(right) but with  form factors. The contribution due to ${\cal O}_0$ is shown without any form factor suppression. This is because we want to show the pure contribution of the operator ${\cal O}_0$ so that any deviation can be interpreted as the presence of higher order effects in $\hat{s}/\Lambda^2$ (as explained in Section~\ref{unitarity}, using a form factor would amount to assuming such higher order corrections).

Finally let us state how the contributions of the different $\g\g ZZ$ operators in eq.~\ref{Lph} can, in principle, be resolved. We have already seen how looking at the final polarization of the $Z$s can be used to distinguish the contribution of ${\cal O}_1^{ZZ}$ and ${\cal O}_2^{ZZ}$ from the other $\g\g ZZ$ operators. Another fact that can be used is that only for the operators ${\cal O}_1^{ZZ}$ and ${\cal O}_3^{ZZ}$ are the amplitudes spherically symmetric. Thus the $\g\g ZZ$  operators in eq.~\ref{Lph} can be divided into four categories: those that contribute mainly to the $Z_L Z_L$ mode and give spherically symmetric amplitudes (only ${\cal O}_1^{ZZ}$), those that contribute mainly to the  $Z_L Z_L$ mode but do not give spherically symmetric amplitudes (only ${\cal O}_2^{ZZ}$), those that contribute mainly to  the $Z_T Z_T$ mode and give spherically symmetric amplitudes (only ${\cal O}_3^{ZZ}$), and those that contribute mainly to  the $Z_T Z_T$ mode but do not give spherically symmetric amplitudes ( ${\cal O}_4^{ZZ}-{\cal O}_6^{ZZ}$). Note that resolving the contributions of the different operators would require higher luminosity than just detecting the presence of QNGCs, but we will not go into the experimental feasibility of such studies.
\section{ LHC signal search strategy}

As explained in Section~\ref{epha} the final state  protons in diffractive processes  are scattered at small angles. To detect such protons very forward detectors have been proposed both for the ATLAS and CMS detectors (see Ref~\cite{AFP}). It has been proposed that such  detectors should be placed at distances of 220 m and 420 m from the interaction point where the distance is along the circular beam line. To give an idea about these distances, at 220 m the beam line curves away from the tangential direction at the interaction point by about 6 meters. The protons lose  a small fraction of their energy in the diffractive process and experience a small deflection from the beam axis. As the deflection is very small, the LHC magnets continue to curve the protons along the beam pipe but they do move away form the beam axis and out of the beam envelope because of the deflection. Thus detectors close to the beam axis (a few millimeters away) would be able to detect the protons. It is also important to note that particles other than protons would never be detected in these detectors as they have a different cyclotron radius. Thus these detectors effectively use the LHC magnets  as a spectrometer.

As the detectors need to be close to the beam axis radiation hardness is a requirement that the detector must fulfill. This along with resolution requirements makes 3D silicon detectors ideal as proton detectors. From the measurement of the position and track direction at the detectors the momentum four vector of the proton can be reconstructed  by inverting the transport of the proton due to  the LHC magnet optics. Thus it is possible to measure the fraction of energy lost by each proton, $\xi_i$ and thus measure the invariant mass of the central system (also called the proton missing mass),
\beq
W=\sqrt{\xi_1 \xi_2 s},
\label{mass}
\eeq
where $s= (14~Ê{\rm TeV})^2$ for the 14 TeV LHC. The 220 m detectors detect protons with smaller deflection, and thus smaller $\xi$, than the 220 m detectors. As  higher invariant masses  would correspond to  higher $\xi$s, the 420 m detector is sensitive in the low mass region whereas the 220 m detector is sensitive in the high mass region. The 220 m detector is  thus crucial for  the kind of study we are doing in this paper where most of the signal contribution comes from events with  high $W$. Mass resolution between 2 GeV and 3 GeV for low energies and about 5-6 GeV for the highest photon energies can be achieved by these detectors~\cite{de Jeneret:2007vi}.

As only events with two intact protons are accepted, the only background processes can be those in which the proton emits a light particle with no electromagnetic or color charge and remains intact. Thus the proton can emit a photon or  a pomeron.  In pomeron fusion processes, also called `Double Pomeron Exchange' (DPE) processes, the pomeron, in general, breaks into fragments. Thus whenever we would write down a process $pp(DPE \to X)pp$, $X$ being a particular final state, it would be implicit that this is an inclusive process where other particles (pomeron fragments) are also present.  These detectors can be used to test if an event is exclusive or inclusive, where by an exclusive event we mean an event where no other particle in addition to the final state particles is  produced. This can be done by matching the invariant mass measured by the proton detectors (using eq.~\ref{mass}) with the invariant mass measured by the central detectors. Also, in exclusive events  the $p_T$ of all the final state particles excluding the protons (which carry very little $p_T$) must add up to nearly zero. Thus if only such exclusive events are accepted  the underlying process can only be a  an exclusive pomeron fusion process, usually called `Central Exclusive Production' (CEP) and the inclusive DPE background can be reduced. Including both the 220 m and 420 m the acceptance range for $\xi_i$ is~Ê\cite{deJeneret:2007vi},
\beq
0.0015<\xi_i<0.15
\label{xiac}
\eeq
Using eq.~\ref{xiac} we see that only   events with  $21$ GeV$<W<2100$ GeV are accepted by the detector.  

A potentially important background contribution is from overlap events. If the signal event is $pp\to pXp$, an overlap event would be defined as the coincidence of an event where the central system $X$ is produced with one or more diffractive events in the same bunch crossing.  Processes like $[p][Xp]$, where one of the protons is produced in an interaction different from the central process (the square brackets grouping the final particles produced in the same scattering process) or processes like $[pp][X]$ and $[p][X][p]$  where both the protons are produced in an interaction different from  the central system, can fake signal events. To reject such background events the forward detectors would be equipped with timing detectors which would have a resolution of the order of 10 ps~\cite{AFP} (note that the LHC bunch length is about 250 ps at 1-sigma). These detectors would be able to reconstruct the vertex position of the two protons assuming they are produced at the same interaction point. This vertex would not match the vertex for the central system, $X$, for the fake events and thus such background events can be rejected. Another way to reduce the overlap background is by matching the net invariant mass of the central system measured by the central detectors with the values obtained by the forward detectors. We  have already discussed how this can be done in the context of testing whether an event is exclusive or not. Overlap backgrounds are  of great importance when the inelastic production cross-section for the central system $X$ is large as is the case for the dijet background to diffractive $H \to bb$ production~\cite{Cox:2007sw}(where the jets are misidentified as $b$-jets) but is of much lesser importance in our case.   Let us now discuss the search strategy for the $pp(\gamma\gamma\to ZZ)pp$ and the $pp(\gamma\gamma\to \g\g)pp$ processes.

\subsection{$pp(\gamma\gamma\to ZZ)pp$ process}

As we do not perform a detailed detector simulation and  wish  to make only  an estimate of the detector level cross-section, we will look at the $p+(ZZ\to 4l)+p$ final state ($l=e,\mu$) that is most free from  experimental complications.  Other final states involving hadronic $Z$ decay  modes may well turn out to  be more sensitive to our observable but ascertaining this would require a more rigorous experimental analysis. 

The most important background process is    $pp(DPE\to ZZ)pp$  in the SM. As mentioned before this is an inclusive process where the final state has pomeron remnants in addition to the $Z$ pair. The DPE background has been computed using the Forward Physics Monte Carlo (FPMC)~\cite{KepkaPhD} which uses the Ingelman-Schlein (IS) model~\cite{Ingelman:1984ns} for inclusive diffraction. In this model  the cross-section of a process like $pp(DPE\to X)pp$ is computed by convoluting the cross-section of the partonic hard process $ij\to X$ ($i,j$ being the partons) with diffractive parton density functions (DPDF) measured at HERA.  The DPDF describes the probability of extracting a certain parton with a given longitudinal momentum  fraction from the proton. The DPDF itself can be expressed as a product of the pomeron flux, measured in other diffractive processes, and the probability of extraction of the parton from the pomeron which breaks into fragments. Whereas the IS model describes diffractive data at the Hadron-Electron Ring Accelerator (HERA) very well its theory prediction for diffractive dijet production at  the Tevatron is larger by a factor of 10~\cite{Albrow:2010yb}. This is usually attributed to the fact that there is some probability for the protons in a DPE process to have a subsequent inelastic interaction which breaks the proton. Thus the cross-section computed in this model must finally be multiplied by the survival probability, $S^2_{DPE}$, which is  the probability that there is no further inelastic interaction between the protons. The maximum value of the survival probability  reported in the literature is about  $S^2_{DPE}=0.06$~\cite{Khoze:2007hx}.

For the $pp(DPE\to ZZ)pp$ process that we are interested in, there are two possible partonic sub-processes, the $qq\to ZZ$ sub-process via $t$-channel quark exchange of a quark $q$ and the $gg \to ZZ$ sub-process that is induced by fermion loops{\footnote {For the computaion of the partonic cross-section of the $gg\to ZZ$ sub-process, the $gg\to h^*\to ZZ$ process has not been included. This contribution is known to interfere negatively and decrease the total cross-section~\cite{Glover} so the background cross-section would have been lower had this contribution been incorporated.}}. The quark component of the DPDF gives the dominant contribution in processes involving diffractive vector boson production like the one we are interested in~\cite{cms}. We apply the following cuts to the signal and background~\cite{deJeneret:2007vi,Kepka:2008yx},
\bea
0.0015<\xi_i<0.15 \label{xicut}
\\
W>300{\rm~GeV}.
\eea
 The first cut above is just the    $\xi$-acceptance cut  for the forward detectors, and the second cut has been applied  mainly to suppress the SM loop background discussed later. As explained in Section~Ê\ref{th}, most of the contribution to the signal cross-section comes from high energies so that the second cut hardly affects the signal. With these cuts  the cross-section we thus obtain for $pp(DPE\to ZZ)pp$  from FPMC   including all $Z$ decay modes is  1.4 fb.  Applying the above cut the signal cross-section (with form factor)  is reduced from 8 fb  to 3 fb in the light Higgs case for $b_i/\Lambda^2=1$ TeV$^{-2}$ and $c_i/\Lambda^2=1$ TeV$^{-4}$, where the   $\xi$-acceptance cut  is responsible for most of the reduction.  
 
As the DPE background discussed above is inclusive unlike the signal  it can be further reduced by  testing if the events are exclusive. This can be done by matching the four-lepton invariant mass measured by the central detector with the invariant mass measured using eq~\ref{mass}. Also the net  $p_T$ of the four leptons must add up to nearly zero (as the protons carry hardly any transverse momentum)  for exclusive events. 

Now let us consider the background contribution  from  the SM loop process $pp(\g\g\to ZZ)pp$. The cross section $\sigma_{\gamma\gamma}(\gamma\gamma\to ZZ)$ has been evaluated in Ref.~\cite{Jikia:1992wm}  to be roughly constant, around  $300$ fb in the range 300 GeV $<\tau<2100$ GeV, where the upper limit is equal to the upper limit obtained by applying the $\xi$-acceptance cut in eq.~\ref{xicut}.  Thus we find for the proton level cross section,
\bea
\sigma &=& S^2_{QED}\int_{2M_Z}^{2100} \frac{dL}{d\tau}\sigma_{\gamma\gamma} d\tau\approx 0.9 \times 300  \int_{2M_Z}^{2100} \frac{dL}{d\tau}d\tau\approx  0.1~ {\rm fb},~~
\label{ineq}
\eea
 where we have imposed the cut $0.0015<\xi_i<0.15$. We find therefore that this background is negligible compared to the signal.

So far in all the instances where we have considered signal or background cross-sections we have not taken into account the branching ratio of the $Z$ to leptons and detector efficiencies. Taking the lepton identification rate to be $90\%$~\cite{tdr} and the proton  detection efficiency in the forward detectors to be 85$\%$~\cite{AFP} we  obtain for the effective cross-section we expect the detectors to measure,
\beq
\sigma_{eff} = 0.56~B(Z\to ll)^2\sigma_{th}, 
\label{seff}
\eeq
where $\sigma_{th}$ is the theoretical cross-section including all $Z$ decay modes.

Finally, let us mention a possible complication that may arise because of the fact that the high energy $Z$s we are considering would be boosted in the lab frame. This would cause the leptons to be collimated along the direction of motion of the parent $Z$. This may give rise to complications in detection of some electron pairs for which the two electrons are not well separated from each other (there is no such issue with muonic decays as muon separation is always efficient for the energies we are considering). We will not try to estimate this effect (see for example for a~\cite{Antoniadis:2009ze}
 more detailed discussion) but in our estimates of  sensitivity in the next section, to give a conservative estimate,  we will provide results considering only muonic decay of the $Z$s in addition to results considering decay of the $Z$s into both electrons and muons. 

\subsection{$pp(\gamma\gamma\to \g\g)pp$ process}

For the $pp(\gamma\gamma\to \g\g)pp$ process we require the presence of  two photons and two protons in the final state. Again the main background is from the $pp(DPE\to \g\g)pp$ process. The $pp(DPE\to \g\g)pp$ cross-section can be estimated from the $pp(DPE\to ZZ)pp$ cross-section by using the fact that the $t$-channel quark exchange sub-process $qq \to ZZ/\g\g$ is the dominant partonic sub-process~\cite{cms}. As the diagrams for this partonic sub-processes in both the cases are the same except for the two  outgoing vertices and external legs in the limit of massless quarks we get,
\bea
 \frac{\sigma(pp(DPE\to \g\g)pp)}{\sigma(pp(DPE\to ZZ)pp)} =(1-4 M_Z^2/\hat{s})^{-1/2} \frac{e^4}{(g/2c_w)^4}\left[\frac{\sum_{u,d,s} Q^4}{\sum_{u,d,s}(v_q^2+a_q^2)^2}\right]
\label{ratio}
\eea
where $v_q$ and $a_q$ are the vector-like and axial vector-like couplings of the quarks to the $Z$ boson and $Q$ is their electric charge. The sum is over the three light quarks for which the  probability of diffractive extraction from the proton is significant and we assume that the diffractive PDFs for three  light quarks are equal. The kinematic factor on the RHS, which is almost unity at high energies, arises because the $Z$-boson unlike the photon is massive. Taking $v_{d,s}=-0.35$, $v_u=0.20$,  $a_u=1/2$, $a_d=-1/2$ and $\hat{s}=(500{\rm ~GeV})^2$, we find this ratio to be 0.3.  We apply the same cuts as in the case of the $pp(\gamma\gamma\to ZZ)pp$ process, that is,
\bea
0.0015<\xi_i<0.15 \label{xicut}\\
W>300 {\rm~GeV} \label{tcut}
\eea
and using the above ratio we obtain this background cross-section to be 0.4 fb.  Again this inclusive DPE background can be further reduced by requiring the two photon invariant mass to match the missing mass evaluated using eq.~\ref{mass}  and by demanding that  $p_{T\g 1}=-p_{T\g 2}$ within experimental resolution.

The SM loop induced $pp(\g\g \to \g\g)$ process in this case has a cross-section that is ${\cal O}(0.01)$ fb and can be ignored~\cite{jikia2, sahin}. An experimental background contribution can come from mis-identification of jets as photons in the $pp(\g\g\to jj)pp$ process. The total inclusive diffractive dijet cross-section at the LHC has been computed by the diffractive Monte Carlo generator DPEMC~\cite{Boonekamp:2003ie} to be 4$\times10^7$ fb~\cite{web}  in the IS model with the cut $E_T>25$ GeV for the jets. Taking the rejection factor of jets for photon identification to be 5000~\cite{tdr} we get a background cross section of about 2 fb which is already smaller than the signal cross-section (with form factor) of 17 fb in the light higgs case for $b_i/\Lambda^2=1$ TeV$^{-2}$ and $c_i/\Lambda^4=1$ TeV$^{-4}$. Further cuts like the $W$-cut in eq.~\ref{tcut} and requiring $p_{T\g 1}=-p_{T\g 2}$ within experimental accuracy  should completely remove this background.  

The effective detector level cross-section is again smaller than the values mentioned so far. Taking the photon identification rate to be 90$\%$~\cite{tdr} and proton detection efficiency in the forward detector to be 85$\%$~\cite{AFP} we get,
\beq
\sigma_{eff} = 0.69~\sigma_{th} 
\label{eff2}
\eeq

Note that for both the $pp(\gamma\gamma\to \g\g)pp$ and $pp(\gamma\gamma\to \g\g)ZZ$ processes we have ignored above the effects of the basic detector  acceptance cuts $p_T>10$ GeV and $\eta<2.5$ for the leptons and photons. As the dominant contribution to the signal cross-section is central and  from high energies, these cuts are expected to have a very small effect.

\begin{table}
\begin{tabular}{ccccccc}
\hline
Couplings&Process&Integrated&$N_{obs}$&$N_{b}$&Confidence&\\
&&Luminosity(fb$^{-1}$)&&&Level(sigma)& \\
\hline
\hline
Case 1: ($850$ GeV)$^{-4}$&$\g\g\to\g\g$&1&12.1&0.3&$>$10\\
Case 1: ($1.8$ TeV)$^{-4}$&$\g\g\to\g\g$&300&133.1&82.8&5.2\\
Case 1: ($850$ GeV)$^{-4}$&$\g\g\to ZZ$&300&7.4(1.9)&1.1(0.3)&4.3(2.1)\\
Case 1: ($750$ GeV)$^{-4}$&$\g\g\to ZZ$&300&11.4(2.8)&1.1(0.3)&6.0(2.9)\\
Case 1:($500$ GeV)$^{-4}$&$\g\g\to ZZ$&300&46.8(11.7)&1.1(0.3)&$>$10(8.1)\\
Case 2:($700$ GeV)$^{-4}$&$\g\g\to ZZ$&300&14.8(3.7)&2.1(0.5)&5.8(3.1)\\
Case 2:($500$ GeV)$^{-4}$&$\g\g\to ZZ$&300&51.3(12.8)&2.1(0.5)&8.2(7.7)\\
Case 3: $\Lambda_T=1.0$ TeV&$\g\g\to\g\g$&1&13.5&0.3&$>$10\\
Case 3: $\Lambda_T=2.4$ TeV&$\g\g\to\g\g$&300&118.2&82.8&3.9\\
Case 3: $\Lambda_T=900$ GeV&$\g\g\to ZZ$&300&12.6(3.2)&1.1(0.3)&6.4(3.6)\\
Case 3: $\Lambda_T=700$ GeV&$\g\g\to ZZ$&300&39.6(9.9)&1.1(0.3)&$>$10(7.1)\\
Case 4:($1.9$ TeV)$^{-2}$&$\g\g\to ZZ$&300&5.3(1.3)&1.1(0.3)&3.3(2.1)\\
Case 4:($2.2$ TeV)$^{-2}$&$\g\g\to ZZ$&300&3.9(1.0)&1.1(0.3)&2.2(1.1)\\
\hline\\
\end{tabular}
\caption{The expected number of observed events $N_{obs}$ and the signal significance for both the processes for different integrated luminosities. The expected number of observed events is evaluated using  $N_{obs}=\sigma^{signal}_{eff}L$ where $L$ is the integrated luminosity, and to evaluate the signal significance the background is assumed to follow a Poisson distribution with mean $N_b=\sigma^{bgr}_{eff}L$. The signal contribution has been evaluated with  a form factor as in eq.~\ref{ff} taking $m=2$, $n=1$ and $\Lambda^{\g\g/ZZ}_f=\Lambda^{\g\g/ZZ}_{UB}$. For the $\g\g \to ZZ$ process the values in the parentheses show the results if only muonic decays of $Z$ are considered. }
\label{sens}
\end{table}

\section{LHC sensitivity to QNGCs }

Using the LHC search strategy for $pp(\gamma\gamma\to \g\g)pp$ and $pp(\gamma\gamma\to \g\g)ZZ$ signals outlined in the previous section we can now report the expected sensitivity of diffractive photon fusion at LHC  to QNGCs.  Table~\ref{sens} shows the expected number of observed events $N_{obs}$ and the signal significance for both the  processes with different integrated luminosities. The expected number of signal, background and observed events are evaluated using,
\bea
 N_{S}=\sigma^{signal}_{eff}{\cal L}_{int}
\nonumber\\
 N_{B}=\sigma^{bgr}_{eff}{\cal L}_{int}
\nonumber\\
N_{obs}=N_S+N_B
\label{stats}
\eea
where  ${\cal L}_{int}$ is the integrated luminosity and $\sigma_{eff}$ is the effective cross-section defined by eq.~\ref{seff} and eq.~\ref{eff2}, after taking into account detector efficiencies. The signal contribution has been evaluated with a form factor as in eq.~\ref{ff} taking $m=2$, $n=1$ and $\Lambda^{\g\g/ZZ}_f=\Lambda^{\g\g/ZZ}_{UB}$. We can simply add the signal and background events  to get the total number of events expected to be observed in eq.~\ref{stats} because the interference with the background is very small. The interference with the DPE background is small because the interference between DPE and photon exchange diffractive processes is in general small and the interference with the SM loop background is small because unlike the signal this background gets most of the contribution from the low-$W$ region. In order to quantify the signal significance we evaluate the probability, $\alpha$, that the background has not fluctuated to give a number of events greater than or equal to $N_{obs}$  assuming that it follows a Poisson distribution with  $N_b$ as its mean. The confidence level expressed as a particular number of sigma deviations is  given by $\sqrt{2}$ erf$^{-1}(\alpha)$ where erf() is the error function. We find the sensitivity for four different physically interesting ways of choosing the relative value of the QNGCs.
\\
\\
\underline{CASE I: $b_i/\Lambda^2=1/\Lambda^2$ , $c_i/\Lambda^4=1/\Lambda^4$}
\\
\\
We find that the $\g\g$ production process is by far the more  promising of the two processes for probing QNGCs. As we can see from  Table~\ref{sens} even with integrated luminosities as low as ${\cal L}_{int}=1$ fb$^{-1}$,couplings as small as $1/(850~{\rm GeV})^{4}$ can be probed with large significance. With high integrated luminosity (300 fb$^{-1}$) couplings as small as $1/(1.8~{\rm TeV})^{4}$ can be detected with more than 5 sigma significance. There are possible cuts that can remove inclusive events as discussed in the previous section, which may substantially reduce the DPE  background. If this is possible the $\g\g \to \g\g$ process can be sensitive to even smaller couplings. Note that a coupling with value $1/(1.8~{\rm TeV})^{4}$ does not necessarily mean that the energy scale of new physics is 1.8 TeV. If dimensionless couplings less than unity or loop factors are present, for instance, the scale of new physics would be lower.

The $ZZ$ production process requires very high integrated luminosity.  For this process we give in addition to the  results assuming $Z$ decays to both electrons and muons, the results considering only the muonic decays in parentheses. For ${\cal L}_{int}= 300$ fb$^{-1}$ as one can see from Table~\ref{sens} the smallest couplings that can be detected with more than 95 $\%$ confidence level are about $1/(850~{\rm GeV})^{4}$. If we require the detection of at least 10 signal events these values are $1/(750~{\rm GeV})^{4}$  and $1/(500~{\rm GeV})^{4}$  considering respectively decays to both electrons and muons and only muonic decays.
\\
\\
\underline{CASE 2: Resolving QNGC contributions to $ZZ$ production from the  contribution due to ${\cal O}_0$}
\\
\\
As mentioned before the dimension-6 operator  ${\cal O}_0$  contributes to the signal through the  $\g\g\to h ^*\to ZZ$ process. The value of this coupling can be obtained from  the $h \to \g\g$ partial width measurement. For this case we consider this contribution to be part of the background and take all the $ b_i/\Lambda^2=1/(850~{\rm GeV})^{4}$and all the $c_i$ equal. We then try to find the smallest QNGC couplings $c_i$ that can be detected. As we want to separate the  ${\cal O}_0$  contribution from higher dimensional contributions, we do not use any form factor for the evaluation of the $\g\g\to h ^*\to ZZ$ cross-section due to this operator  as using the form factor is equivalent to including higher dimensional corrections (see Section~\ref{unitarity}). We find that the smallest couplings that can be detected for this case with 300 fb$^{-1}$ data to be $1/(700~{\rm GeV})^{4}$ ($1/(500~{\rm GeV})^{4}$) considering  $Z$-decays to both electrons and muons (to only muons). 
\\
\\
\underline{CASE 3: Graviton exchange in extra-dimensional model}
\\
\\
For this case we assume that the  QNGCs arise from the effective dimension-8 operator due to virtual graviton exchange in extra-dimensional theories described in  Section~\ref{extrad}. The relative couplings of the QNGCs are thus fixed by expanding the operator in eq.~\ref{vg} and the only adjustable parameter is $\Lambda_T$. 


As one can see from Table~\ref{sens} we find that the $\g\g \to \g\g$ process can detect this operator with only 1 fb$^{-1}$ data for $\Lambda_T=1.0$ TeV. For high luminosities (300 fb$^{-1}$) the maximum value of $\Lambda_T$ that can be probed by this process in the DPE background is about $\Lambda_T=2.4$ TeV. Note that our results differ from and are less optimistic than the results of Sahin et al~\cite{sahin} who do not consider the DPE background and more importantly use a far less restrictive $\xi$-acceptance cut. As explained after eq.~\ref{xiac} because of the $\xi<0.15$ acceptance cut only events with $W<2.1$ TeV are accepted. In Ref.~\cite{sahin}, on the other hand, events with $\xi$ as high as 0.5 are accepted which corresponds to $W$ as high as 7 TeV and most of the contribution to their signal comes from the high $\xi$ events; protons with $\xi>0.15$ can, however, not be detected by the forward detectors~\cite{de Jeneret:2007vi,Kepka:2008yx}.  For the $\g\g \to ZZ$ process $\Lambda_T$ as high as 900 GeV (700 GeV) can be probed with 300 fb$^{-1}$ data considering  $Z$-decays to both electrons and muons (to only muons). As already mentioned in Section~\ref{extrad}  the $gg\g\g/ggZZ$ operators ($g$ being a gluon) that arise from expanding the operator in eq.~\ref{vg} are expected to give a contribution to the exclusive $pp\g\g/ppZZ$ final states via central exclusive pomeron fusion but this contribution is expected to be negligible relative to the diffractive photon fusion contribution. 

Establishing the presence of QNGCs would give very important complementary evidence for virtual graviton exchange because it is possible in this case to uniquely trace back to the underlying dimension 8 operator involved. Our final sensitivity results show, however, that for the particular diffractive processes we have studied for probing QNGCs the largest $\Lambda_T$ that can possibly be probed (2.4 TeV) has already been ruled out by dijet constraints from the 36 pb$^{-1}$ CMS data in Ref.~\cite{Franceschini:2011wr} where the constraint $\Lambda_T>3.8$ TeV has been derived.   Thus diffractive photon fusion will not be able to probe $\Lambda_T$ values still allowed by experimental data.
\\
\\
\underline{CASE 4: Higgsless case}
\\
\\
As we discussed in Section~\ref{higgsless} in the higgsless case we expect only the following  two operators to be important,
\beq
{\cal L}^{Higgsless}_{QNGC} = \frac{(g/2c_w)^2a_1^{hl}}{\Lambda^2} F_{\mu \nu}F^{\mu \nu} Z_\rho Z^\rho+ \frac{(g/2c_w)^2 a_2^{hl}}{\Lambda^2}F_{\mu \nu}F^{\mu \rho} Z_\rho Z^\nu.
\eeq 
 We take $a_1^{hl}=a_2^{hl}$ in eq.~\ref{laghl} and find that the $pp(\g\g \to ZZ)pp$ process is sensitive up to couplings as small as  $1/(1.9$ TeV)$^2$   if we require more than 95$\%$ confidence level, a huge improvement over existing limits. Our sensitivity estimates agree  well  with  those obtained by Royon et al. in Ref.~Ê\cite{Kepka:2008yx}, once we translate to their convention for parametrization of these couplings. 

Higgsless models are usually associated with strong electroweak symmetry breaking (EWSB) scenarios. The operator coefficients in such theories can be estimated by Naive Dimensional Analysis (NDA) (see Refs~\cite{Manohar:1983md,Giudice:2007fh}).  In our case, using NDA, we find $a_i^{hl}/\Lambda^2=e^2/(16 \pi^2 \Lambda_s^2)$, $\Lambda_s$ being the scale of the strongly coupled sector.\footnote{Note that the $ZZZZ$ coupling which appears at the dimension 4 level in the chiral lagrangian, from operators like $c~({\rm Tr}[(D_\mu \Sigma)^\dagger D^\mu \Sigma])^2$, is less suppressed ($c \sim v^2/\Lambda_s^2\approx 1/(16 \pi^2)$) than other QNGCs.} The above mentioned estimates tell us that our process is sensitive to $\Lambda_s< 100$ GeV. As a strong sector at such low energies is already ruled out by experiments, our process, unfortunately, cannot probe realistic scales for strong EWSB.
\section{Conclusions}
We have listed all possible operators contributing, at the lowest order,  to Quartic Neutral Gauge Couplings, quartic gauge couplings involving only the photon and the $Z$ boson and have studied the sensitivity of measurement of these couplings in diffractive photon fusion processes at the LHC. These couplings are interesting because the lowest order contribution they receive is from dimension 8 operators in scenarios with a light Higgs and,   in higgsless scenarios,  from  dimension 6 operators  (with the exception of the $ZZZZ$ coupling which receives dimension 4 contributions in this scenario but we have not focussed on this coupling  in this work in any case). Thus new physics processes which do not contribute through operators of the lowest possible dimension can be probed by measuring these couplings. One specific example that we have considered is virtual graviton exchange in extra dimensional theories where the lowest dimension operators generated are of dimension 8, and these include operators contributing to QNGCs.

 Thus measurement of QNGCs in any experimental process would be interesting, but in this work we have studied  their measurement in diffractive photon fusion processes like $pp(\g\g\to \g\g)pp$ and $pp(\g\g\to ZZ)pp$. The protons in these processes remain intact  and scatter diffractively with very small scattering angles. These can be detected by very forward proton detectors that have been proposed for both the ATLAS and CMS experiments. As  we argue the detection of the two $\g/Z$s in the central detectors along with the detection of the protons in these forward detectors  would indicate the existence of QNGCs like the  $\g\g\g\g$ and $\g\g ZZ$ couplings,    as this is the only feasible new physics possibility that can lead to such a final state. The only other possibility  is $pp(CEP \to ZZ)pp$, where CEP stands for Central Exclusive Production, is a  process  that takes place when pomerons fuse exclusively (that is without breaking into fragments) to give the $ZZ/\g\g$ final state. Such processes are, however, expected to have a much smaller cross-section when compared to photon fusion processes. To calculate the cross-section for the $pp(\g\g\to \g\g)pp$ and $pp(\g\g\to ZZ)pp$ processes we convolute the cross-section of the $\g\g \to \g\g/ZZ$ sub-process with the $\g\g$ luminosity function obtained using the Equivalent Photon Approximation. The amplitude of the $\g\g \to \g\g/ZZ$ sub-process grows with energy because of the non-renormalizable couplings involved and we unitarize this using appropriate form factors. We have argued that  our final sensitivity results for $\Lambda$ will not change much for a different choice of form factor than ours. 

Before we summarize our results on the sensitivities, note that QNGCs are very weakly constrained by existing data. Whereas no constraints exist on $\g\g\g\g$ couplings, the $\g\g ZZ$ couplings are constrained by direct search results from LEP to be smaller than about $1/(100~$ GeV$)^4$ ($1/(50~$ GeV$)^2$) and by precision measurements to be smaller than about $1/(270~$GeV$)^4$ ($1/(420~$GeV$)^2$) in the light Higgs (higgsless) case. We have found in this study that diffractive photon fusion at LHC can improve these sensitivities by many orders of magnitude for the $\g\g ZZ$ coupling, and can probe couplings as small as $1/(850$ GeV)$^{4}$ (1/(1.9 TeV)$^{2}$) with 300 fb$^{-1}$ integrated luminosity for the light Higgs case (higgsless case). We find, however,  using an NDA estimate, that the values  in the higgsless case correspond to a scale lower than 100 GeV for the strong sector which is already excluded by experiments.  The $\g\g\g\g$ coupling can be probed even more sensitively and values as small as $1/(1.8$ TeV)$^{4}$ can be measured with the same integrated luminosity for the light Higgs case. For the specific case of virtual graviton exchange in theories with large extra dimensions we find that the highest scale that can be possibly probed (about $\Lambda_T=2.4$ TeV by the $pp(\g\g \to \g\g)pp$ process with 300 fb$^{-1}$ data) has, unfortunately,  already been ruled out by the latest constraint from CMS dijet data which puts the bound $\Lambda_T>3.8$ TeV.

\noindent
\begin{bf}
Acknowledgments\end{bf}: We would like to  thank, first of all, C. Grojean and J.D. Wells for  valuable comments at all stages of the project.  We thank O. Kepka and C. Royon for providing us with the Double Pomeron Exchange background and for clarifying issues about diffractive processes.  We also thank S. Rychkov for  discussion on the NDA estimates in the higgsless case. Finally, we are grateful to  N. Desai, A. Khmelnitskiy, D. Pappadopulo and S. Prestel   for helpful conversations.
This work is supported in part by  the European
Commission under the contract ERC advanced grant 226371 `MassTeV'.
\\
\\
 {\Large   {\it Appendix A: $\g\g\g Z,~\g ZZZ$ and $ZZZZ$ couplings}}
\newline

We first consider the operators in  the light higgs case in eq.~\ref{L8} when written in terms of the fields $A$ and $Z$  give rise to $\g\g\g\g$, $\g\g ZZ$, $\g\g\g Z$,  $\g ZZZ$ and $ZZZZ$ couplings. We already wrote the Lagrangian for the $\g\g\g\g$ and $\g\g ZZ$ couplings in eq.~\ref{Lph}. In this Appendix we will present the $\g\g\g Z,~\g ZZZ$ and $ZZZZ$ couplings. Using eq.~\ref{transform2} we get from eq.~\ref{L8},
\bea
{\cal L}^{\g\g\g Z}_{QNGC} &=&  \frac{a_1^{\g\g \g Z}}{\Lambda^4} F_{\mu \nu}F^{\mu \nu} F_{\rho \sigma}Z^{\rho \sigma} +\frac{a_2^{\g\g \g Z}}{\Lambda^4} F_{\mu \nu}F^{\mu \rho} F_{\rho \sigma}Z^{\sigma \nu}
\nonumber\\
{\cal L}^{\g ZZZ}_{QNGC} &=& \frac{a_1^{\g ZZZ}}{\Lambda^4}\frac{M_Z^2}{2} F_{\mu \nu}Z^{\mu \nu} Z_\rho Z^\rho+ \frac{a_2^{\g ZZZ}}{\Lambda^4}\frac{M_Z^2}{2}F_{\mu \nu}Z^{\mu \rho} Z_\rho Z^\nu+ \frac{a_3^{\g ZZZ}}{\Lambda^4} F_{\mu \nu}Z^{\mu \nu} Z_{\rho \sigma}Z^{\rho \sigma}
\nonumber\\&&
+ \frac{a_4^{\g ZZZ}}{\Lambda^4} F_{\mu \nu}Z^{\mu \rho} Z_{\rho \sigma}Z^{\sigma \nu}
\nonumber\\
{\cal L}^{ZZZZ}_{QNGC} &=&\frac{a_1^{ZZZZ}}{\Lambda^4}\frac{M_Z^4}{4} Z_{\mu }Z^{\mu } Z_\rho Z^\rho + \frac{a_2^{ZZZZ}}{\Lambda^4}\frac{M_Z^2}{2} Z_{\mu \nu}Z^{\mu \nu} Z_\rho Z^\rho
+ \frac{a_3^{ZZZZ}}{\Lambda^4}\frac{M_Z^2}{2}Z_{\mu \nu}Z^{\mu \rho} Z_\rho Z^\nu
\nonumber\\&&
+ \frac{a_4^{ZZZZ}}{\Lambda^4} Z_{\mu \nu}Z^{\mu \nu} Z_{\rho \sigma}Z^{\rho \sigma}
+ \frac{a_5^{ZZZZ}}{\Lambda^4} Z_{\mu \nu}Z^{\mu \rho} Z_{\rho \sigma}Z^{\sigma \nu}
\label{Lphap}
\eea
where for the $\g\g\g Z$ couplings we get,
\bea
a_1^{\g\g\g Z}&=&-4s_w  c_w^3 c_{8}+ 4 s_w^3 c_w (c_{9}+c_{10})+(2 s_w c^3_{w} -2s^3_w c_w)(c_{11}+c_{12})\nonumber\\
a_2^{\g\g\g Z}&=&-4s_w  c_w^3 c_{13}+ 4 s_w^3 c_w (c_{14}+c_{15})+(2 s_w c^3_{w} -2s^3_w c_w)(c_{16}+c_{17}),
\eea
for the $\g ZZZ$ couplings we get,
\bea
a_1^{\g ZZZ}&=&-2 s_w c_w c_3 +2 s_w c_w c_4-(c_w^2-s_w^2) c_5\nonumber\\
a_2^{\g ZZZ}&=&-2 s_w c_w c_6 +2 s_w c_w c_7\nonumber\\
a_3^{\g ZZZ}&=&-4s_w^3  c_w c_{8}+ 4 s_w c_w^3 (c_{9}+c_{10})+(2 s^3_w c_{w} -2s_w c^3_w)(c_{11}+c_{12})\nonumber\\
a_4^{\g ZZZ}&=&-4s_w^3  c_w c_{13}+ 4 s_w c_w^3 (c_{14}+c_{15})+(2 s^3_w c_{w} -2s_w c^3_w)(c_{16}+c_{17}).
\eea
and for the $ZZZZ$ couplings we get,
\bea
a_1^{ZZZZ}&=&c_1+c_2\nonumber\\
a_2^{ZZZZ}&=&s_w^2 c_3 +c_w^2 c_4+c_w s_w c_5\nonumber\\
a_3^{ZZZZ}&=&s_w^2 c_6 +c_w^2 c_7\nonumber\\
a_4^{ZZZZ}&=&s_w^4 c_{8}+ c_w^4(c_{9}+ c_{10})+c_w^2 s_w^2 (c_{11}+c_{12})\nonumber\\
a_5^{ZZZZ}&=&s_w^4 c_{13}+ c_w^4(c_{14}+ c_{15})+c_w^2 s_w^2 (c_{16}+c_{17}).
\eea

As explained in Section~Ê\ref{higgsless},  any U(1)$_{em}$ invariant operator, constructed  using $Z_\mu$ and $F_{\mu \nu}$ fields,  is an allowed operator in the higgsless case.  Thus, for the higgsless case we will get the same operators as above but now ${\cal O}_1^{ZZZZ}$ would arise from dimension 4 operators while ${\cal O}_1^{\g ZZZ}$, ${\cal O}_2^{\g ZZZ}$, ${\cal O}_2^{ZZZZ}$ and ${\cal O}_3^{ZZZZ}$ would arise from dimension 6 operators. 
\\
\\
\\
 {\Large   {\it Appendix B: Derivation of unitarity relation}}
\newline

In this appendix we derive the expression for the unitarity bound for the processes $\g\g\to \g\g$ and $\g\g\to ZZ$. Applying the optical theorem to the $\g\g\to \g\g$ process  tells us,
\bea
\frac{\Im({\cal M}(\gamma_1 \gamma_2 \rightarrow \gamma_1 \gamma_2))}{s}&=&\sigma(\gamma_1 \gamma_2\rightarrow {\rm everything})
\nonumber\\
&=&\sigma(\gamma_1 \gamma_2\rightarrow \gamma(\epsilon_1) \gamma(\epsilon_2))+\sum_{\epsilon_3,\epsilon_4}\sigma(\gamma_1 \gamma_2\rightarrow Z(\epsilon_3)Z(\epsilon_4))
\nonumber\\
&&+\Delta
\label{optical}
\eea
where $\gamma_i$ denotes $\gamma(k_i,\epsilon_i)$ and $\Delta$ is a positive number that accounts for all the other contributions to the RHS of eq.~\ref{optical} and the cross section for the $\gamma \gamma\rightarrow VV$ process is given by,
\beq
\sigma=\frac{\beta_W}{64\pi^2 s} \int d\Omega_{{\rm CM}}~|{\cal M}(\gamma_1 \gamma_2\rightarrow VV)|^2.
\label{cross}
\eeq
The amplitude can be expanded into partial waves as follows,
\bea
{\cal M}(\gamma_1 \gamma_2 \rightarrow \gamma_1 \gamma_2)&=&16 \pi \sum_J (2J+1)b_J P_J(\cos \theta) \nonumber\\
{\cal M}(\gamma_1 \gamma_2\rightarrow ZZ)&=&16 \pi \sum_J (2J+1)a_J P_J(\cos \theta).
\label{partial}
\eea
where $\beta_V=\sqrt{1-\frac{4M_V^2}{s}}$. For the forward scattering in the LHS of eq.~\ref{optical}, we must put $\theta=0$. Using eq.~\ref{optical},~\ref{partial} and~\ref{cross} and the following  property of Legendre polynomials, 
\beq
\int_{-1}^{1} P_m (x) P_n(x) = \frac{2}{2n+1} \delta_{mn},
\eeq
gives,
\beq
(\Im(b_l))^2 -\Im(b_l)+\sum_{\epsilon_3,\epsilon_4}({\rm Re}(b_l))^2+ \beta_W\sum_{\epsilon_3,\epsilon_4}|a_l|^2 +\delta_l=0.
\label{quad}
\eeq
The first two terms in eq.~Ê\ref{quad} should be evaluated taking the initial polarizations to be exactly same as the final polarizations, and $\delta_l$ is the positive contribution from every other source. Eq.~\ref{quad} is a quadratic equation for $\Im(b_l)$. The equation must have real roots and thus must have a positive discriminant. This gives the condition,
\beq
( {\rm Re}(b_l))^2+\beta \sum_{\epsilon_3,\epsilon_4} |a_l|^2+\delta_l<\frac{1}{4}.
\label{UC}
\eeq
\\
\newline
 {\Large   {\it Appendix C: Kinematic bound on photon virtuality}}
\newline

First let us derive the kinematical limits on $q_i^2$. From conservation laws we must have $q=(E-E'_i,\vec{p}-\vec{p}'_i)$. Substituting $|\vec{p}'_i|=\sqrt{E_i^2-m^2}$, $m$ being the mass of the particle emitting the photon, we obtain for $m\ll E'_i$,
\beq
q_i^2=-4E E'_i \sin^2\frac{\theta_i}{2}-\frac{m^2 \omega_i^2} {EE_i}\cos \theta_i.
\eeq
Here $\theta_i$ is the angle between $\vec{p}$ and $\vec{p}'_i$. In the expression above the first term dominates. As most of the contribution to the amplitude comes from the small $|q_i^2|$ region, ignoring the second term above we see that we must have small $\theta$. We thus obtain the following kinematical bound on $q_i^2$,
\beq
q_i^2 <- \frac{m^2 \omega_i^2} {E(E-\omega_i)}.
\eeq

\end{document}